\documentclass{article}

\usepackage{amsmath,amsfonts,graphicx,amssymb,natbib,multirow,hyperref}

\title{Physical drivers of ocean wave attenuation in the marginal ice zone}

\author{Fabien Montiel$^{1}$, Alison L.\ Kohout$^{2}$ and Lettie A.\ Roach$^{3}$}

\date{$^{1}$University of Otago, Dunedin, New Zealand\\
$^{2}$National Institute for Water and Atmospheric Research, Christchurch, New Zealand\\
$^{3}$University of Washington, Seattle, USA
}

\begin{document}

\maketitle

\begin{abstract}
	Despite a recent resurgence of observational studies attempting to quantify the ice-induced attenuation of ocean waves in polar oceans, the physical processes governing this wave attenuation phenomenon are still poorly understood. 
Most analyses have attempted to relate the spatial rate of wave attenuation to wave frequency, but have not considered how this relationship depends on ice, wave and atmospheric conditions. 
An in-depth analysis of the wave-buoy data collected during the 2017 PIPERS programme in the Ross Sea is conducted. 
Standard techniques are used to estimate the spatial rate of wave attenuation $\alpha$ and the influence of a number of potential physical drivers on its dependence on wave period $T$ is investigated. 
A power-law is shown to consistently describe the $\alpha(T)$ relationship, in line with other recent analyses. 
The two parameters describing this relationship are found to depend significantly on sea ice concentration, mean wave period and wind direction, however. 
Looking at cross-correlations between these physical drivers, three regimes of ice-induced wave attenuation are identified, which characterise different ice, wave and wind conditions, and very possibly different processes causing this observed attenuation. 
This analysis suggests that parametrisations of ice-induced wave decay in spectral wave models should be piecewise, so their dependence on local ice, wave and wind conditions is described in some way.
\end{abstract}

\section{Introduction}

The Southern Ocean has experienced the most intense changes in ocean wave activity resulting from the observed increase in extreme weather events caused by climate change \citep{young_etal11,young_ribal19}.
As the frequency and magnitude of large wave events continue to grow under projected climate change scenarios \citep{meucci_etal20}, it has been conjectured that sea ice morphology will be increasingly affected by the climatology of ocean waves at high latitudes \citep{Kohout_etal14}. 
Both the growth and melt seasonal sea ice cycles are impacted by local wave conditions, e.g.\ as pancake ice is more likely to form in a wavy sea \citep{nose_etal21}, while wave-induced sea ice breakup magnifies the thawing rate of the ice cover in spring and summer \citep{bennetts_etal17}.
The outer band of the ice cover where ocean waves and sea ice dynamics are strongly coupled is referred to as the marginal ice zone (MIZ).

In turn, the extent to which sea ice dynamics respond to changes in wave conditions depends significantly on how quickly wave energy attenuates with distance from the ice edge in the MIZ.
This spatial rate of ice-induced wave attenuation is understood to be governed by (i) a conservative multiple scattering process \citep{montiel_etal16} and (ii) a collection of dissipative physical processes, e.g.\ under-ice turbulence and overwash \citep[e.g.][]{voermans_etal19,nelli_etal20}.
Despite a recent explosion of research in the modelling and observation of these phenomena \citep{squire20}, much remains to be understood with regards to their relative importance as a function of ice and wave conditions. 
Using spectral wave modeling, \cite{ardhuin_etal20} argue that dissipation caused by anelastic ice deformations dominates over other processes in explaining the observed attenuation of significant wave height in the Ross Sea winter MIZ.
In particular, these authors suggest attenuation caused by scattering in a broken ice field is negligible, as long-crested, low-frequency waves travel nearly unaffected by the presence of sea ice under such conditions, confirming similar earlier findings in the Arctic Ocean \citep{ardhuin_etal16}.
However, the spatial rate of wave energy attenuation, usually referred to as the attenuation coefficient and denoted by $\alpha(T)$, depends on the wave period $T$, so that different mechanisms are likely to govern attenuation in different spectral bands of the wave spectrum, with scattering only important at short wave periods.

Field work conducted in the 1970's and 1980's by the Scott Polar Research Institute in the Arctic Ocean showed that the wave energy spectrum $E(T)$ decays exponentially with distance from the ice edge, $x$ say, i.e.\ $E(T) \propto \exp(-\alpha(T)x)$ \citep{squire_moore80,wadhams_etal88}.
These observations consistently show $\alpha$ decreases with increasing $T$ values over a spectral range covering mid to long wave periods, i.e.\ $T\gtrapprox10$\,s. 
At shorter periods, the so-called rollover effect is observed in the data, whereby $\alpha$ reaches a maximum at a wave period typically between 5 and 10\,s and then decreases with decreasing $T$ values.
Although the decreasing trend of $\alpha(T)$ can be understood in the context of both scattering theory \citep{montiel_etal16} and effective viscous media theory \citep{mosig_etal15,meylan_etal18}, the rollover effect is more puzzling and only speculative explanations of the physical mechanism describing this feature, e.g.\ local wind wave generation or nonlinear wave-wave interaction \citep{li_etal17}, have been proposed.
That is until the recent study by \cite{thomson_etal21}, who demonstrated using simulated wave spectra that a low signal-to-noise ratio at low wave periods results in a negative bias in the estimated attenuation coefficient in this wave period range.
This suggests that the rollover is an artefact of the instrument noise and subsequent data analysis as opposed to an emergent property of physical processes.
This finding also suggests the relationship between $\alpha$ and wave period $T$ can be parametrised simply.

A power-law relationship of the form $\alpha \propto T^{-\beta}$, with $\beta>0$, has been proposed to describe the spectral dependence of the attenuation coefficient.
A number of observational studies found this relationship fits attenuation data reasonably well in both the Southern and Arctic Oceans \citep[see][for an extensive survey]{meylan_etal18,rogers_etal21}.
What is more, the power parameter $\beta$ seems to be consistently bounded between 2 and 4, and to hold for a range of ice conditions, including pancake ice, array of broken ice floes or continuous pack ice.
Two key questions naturally arise from these observations: (i) how does this simple relationship emerge from the underlying physical processes governing ice-induced wave attenuation and (ii) how does the parameter $\beta$ depend on observable quantities describing the state of the wave/ice system (e.g.\ ice concentration, thickness, wave height, etc)?
Although question (i) largely remains unanswered,
\cite{meylan_etal18} showed how such a power-law emerges from the dispersion relation of several homogenised viscoelastic layer models in some asymptotic limits, noting the large variance of the power $\beta$ (between 3 and 11).
They also show that a desired integer $\beta$ value can be obtained within this modelling framework by choosing an energy-loss mechanism adequately, which potentially opens the door to understand the physics of ice-induced wave attenuation better.
Very little has been done to address question (ii), which is the focus of our investigation.
In particular, we seek to explore the extent to which a number of observable physical drivers affect the power-law relationship between $\alpha$ and $T$.

The dataset collected during the Polynyas, Ice Production, and seasonal Evolution in the Ross Sea (PIPERS) expedition is analysed here.
Fourteen drifting wave buoys were deployed in the 2017 austral Autumn (April--June), with the goal of better understanding wave attenuation in the MIZ during the ice growth season.
Buoys remained active for up to 6 weeks, therefore making this dataset, to the authors' knowledge, the largest one ever in terms of the number of wave spectra recorded in the MIZ.
\cite{kohout_etal20} gave a comprehensive description of the PIPERS deployment and analysed the attenuation rate of the frequency-integrated significant wave height.
They found that their estimated attenuation rates are positively correlated to ice concentration and negatively correlated to peak wave period. 
The spectral dependence of the attenuation coefficient has also been quantified in a subsequent analysis of a subset of the full PIPERS dataset \citep{rogers_etal21}. 
In that study, the authors estimated the attenuation coefficient using the model inversion technique introduced by \cite{rogers_etal16}.
Specifically, a regional configuration of Wavewatch III\textsuperscript{\textregistered} (WW3), using a nested grid in the region of the deployment and boundary forcing from a global model hindcast, was used to simulate the transport of phase-averaged wave spectra during the period of interest.
The inversion procedure then consists of optimising the attenuation coefficient for each frequency bin that best fits the spectrum recorded by a wave buoy, assuming it is constant between the ice edge and the buoy.
This method differs from the more traditional approach to estimate the attenuation coefficient using exponential decay curves fitted through the spectra of buoy pairs.
There are advantages and drawbacks with each method. For instance, the model inversion method makes several non-validated assumptions related to the form of the source terms in WW3 (e.g.\ linear scaling by ice concentration), while the buoy pair fitting method assumes stationary wave conditions for each attenuation estimate and requires knowledge of direction of wave propagation, which can be difficult to estimate.
Here, we use the latter of the two methods. 
Efforts are made throughout the paper to discuss the impact of the assumptions on the results.

Our goal is to quantify the influence of wave, ice and atmospheric observable physical drivers on the attenuation coefficient.
These drivers are ice concentration, significant wave height and peak wave period, wind speed and wind direction.
The variable conditions during which the PIPERS dataset was collected allows us to explore meaningfully the impact of these quantities on the attenuation coefficient and its spectral dependence.
\cite{rogers_etal21} also considered the influence of ice concentration, ice thickness, distance to the ice edge and significant wave height.
They found a strong positive correlation between $\alpha(T)$ and thickness, and also distance to the ice edge (the too latter quantities clearly not being independent), a weak positive correlation between $\alpha(T)$ and concentration, and a strong negative correlation between $\alpha(T)$ and significant wave height.
They also fitted several power-law curves through the $\alpha(T)$ curves, but did not investigate the effect of physical drivers on the power parameter $\beta$.
Our analysis therefore extends that of \cite{rogers_etal21} as we seek to characterise the effect of potential physical drivers (including some not considered by these authors, e.g.\ wind) on the power-law relationship describing the dependence of $\alpha$ on $T$, and identify possible multiple regimes of wave attenuation in the MIZ depending on wave, ice and wind conditions.
 
\section{Wave and ice observations}\label{sec:data}

During the PIPERS voyage, 14 drifting WIIOS buoys were deployed in the ice-covered Ross Sea to measure ocean wave spectra in the MIZ under a growing Autumn ice regime.
Each buoy captured 11 minutes of wave data every 15 minutes, and returned an average power spectral density (PSD) $S(f)$ (measured in m$^2$/Hz) every quarter of an hour through spectral analysis \citep[see][]{kohout_etal20}.
Wave direction was not recorded by the WIIOS buoys, so the PSD is only a function of wave frequency $f$ in each 15-minute time period.
The time evolution of the buoy positions and sea ice concentration \citep[retrieved from NSIDC passive microwave satellite data, see][and \textsection\ref{sec:an}.\ref{sec:an-conc}]{peng_etal17,nsidc_conc_data} in the deployment area are shown in the animation provided in the supplementary material.

\begin{figure}[t]
    \centering
    \includegraphics[width=1\textwidth]{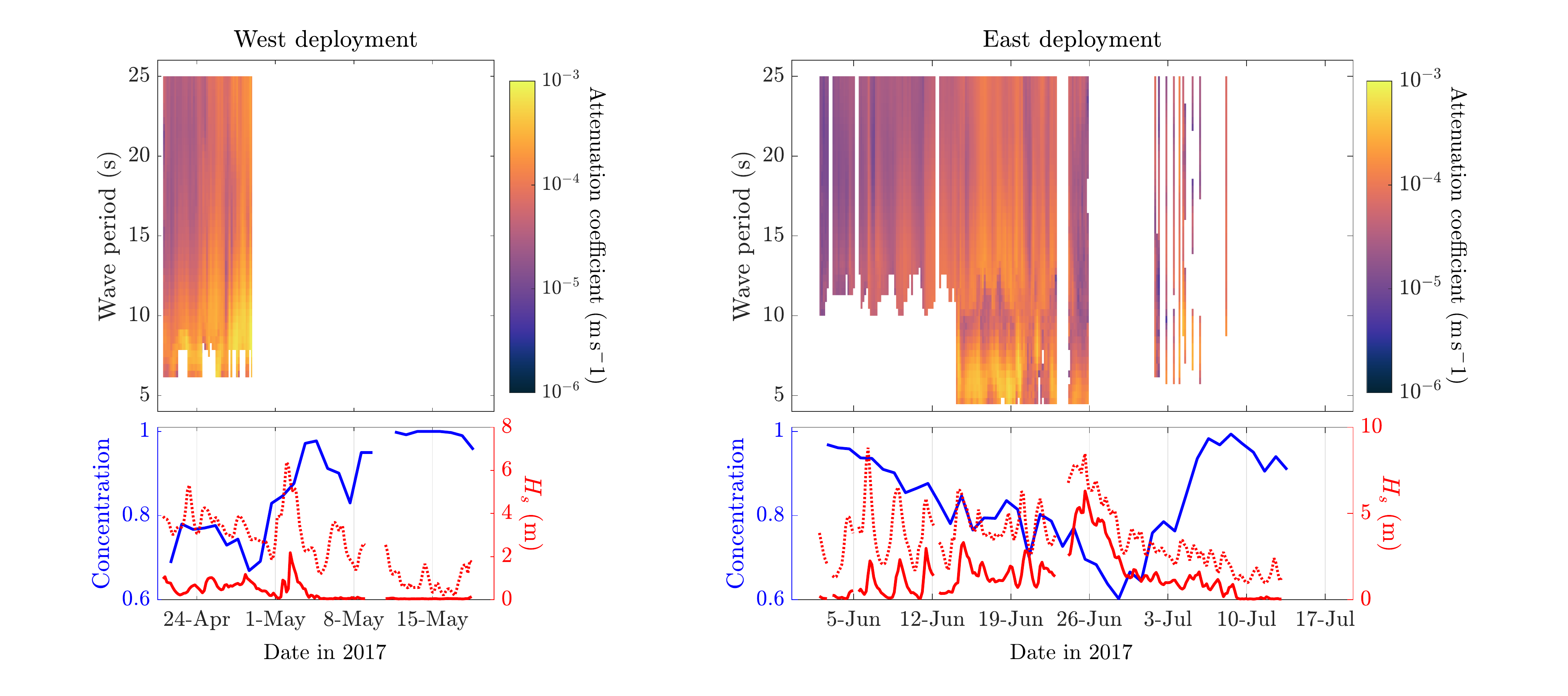}
    \caption{(Top panels) Time-period spectrograms of the attenuation coefficient averaged over all available buoy pairs on successive 4-hour time windows. 
    (Bottom panels) Time series of the average ice concentration (blue solid curve), and average SWHs measured by the active buoys in the MIZ (solid red curves) and by WW3 hindcast data in the open ocean, directly north of the region where the buoys are located (red dotted curve). 
    Left and right pairs of panels correspond to west and east deployment data, respectively.}
    \label{fig1}
\end{figure}

Four buoys were first deployed along a 100\,km-long southward transect as the ship entered the MIZ on 21 and 22 April 2017, in a sector approximately located at 69--70$^{\circ}$S and 171.5--172$^{\circ}$E. 
During this deployment (later referred to as west deployment), two of the buoys only provided data for a few days, while the other two survived for several weeks. 
Figure \ref{fig1} (bottom left panel) shows the evolution of the average ice concentration around the buoys (blue line) and the average significant wave height (SWH) $H_s$ recorded by the buoys (solid red line) during this deployment.
Several moderate wave events were recorded in the early days of the deployment with average SWH peaking at 1--2\,m, while ice concentration was 0.6--0.7. 
From 4 May onward, a significant freeze up event consolidated the ice cover in the sector where the two remaining buoys were located. 
In the following weeks, the buoys recorded low wave activity likely due to the growing ice attenuating south-travelling waves, despite the fact the two buoys drifted north during that period (see supplementary material). 
This is consistent with WW3 hindcast data (multigrid global run using version 4.15 with NCEP wind and ice concentration forcing; see dotted red line in figure \ref{fig1}), which show low to moderate wave conditions ($H_s<4$\,m) consistently in the Western Ross Sea during that period.
We note that \cite{rogers_etal21} did not consider wave data from the west deployment in their analysis.

As the ship left the ice, 10 more buoys were deployed on 2--4 June 2017 along a 250\,km-long transect approximately located at 67--69.3$^{\circ}$S and 184$^{\circ}$E (later referred to as the east deployment). 
The GPS on one buoy failed prior to deployment and so its record is excluded from the analyzed data set.
Six of the nine buoys survived at least three weeks and four survived over a month.
Average ice concentration and significant wave height during the east experiment are shown in Figure \ref{fig1} (bottom right panel).
The average concentration steadily decreased from approximately 1 to 0.6 during the month of June, which can be explained by the buoys drifting north while the ice experienced a significant retreat event (see supplementary material).
Several large wave events were recorded in that period with average SWH between 2 and 4\,m, and one wave event late June reaching an average SWH of approximately 6.3\,m. 
These can be associated with seven very large wave events in June captured by the WW3 hindcast data with SWH 6--9\,m.
It is possible that the storms causing these wave events, and possibly the waves themselves, are partially responsible for the observed ice retreat.
From late June onward, sea ice rapidly grew coinciding with calmer wave conditions.

In situ visual observations of ice thickness, concentration, ice type and floe size were performed along the ship track following the Antarctic Sea Ice Processes and Climate (ASPeCT) protocol.
Visual estimates and ice cores, where possible, were collected to measure the thickness of each ice floe hosting a wave buoy.

\section{Analysis}\label{sec:an}

\subsection{Spectral filtering}
\label{sec:an-filt}

As shown in \textsection \ref{sec:data}, wave buoys recorded spectra that span a wide range of wave conditions, including large wave events and very calm seas. 
\cite{thomson_etal21} showed that low-energy spectral components recorded by accelerometers, as is the case for the present buoys, are easily polluted by a non-uniform noise level which depends on frequency as $f^{-4}$.
Although open ocean spectra commonly have a high-frequency spectral tail following this same power-law dependence, wave spectra measured in ice-covered oceans often exhibit a steeper tail, so that there exists a frequency $f_{cut}$, such that for $f>f_{cut}$ the noise level in the recorded spectra is larger than the true PSD.
Although the high noise-to-signal ratio at high frequencies has little effect on frequency-averaged spectral metrics, e.g.\ the SWH, \cite{thomson_etal21} showed that it affects measurements of frequency-dependent wave attenuation in ice-covered seas by creating a spurious rollover effect, whereby the rate of exponential attenuation decreases with frequency as $f>f_{cut}$.

\begin{figure}[t]
    \centering
    \includegraphics[width=1\textwidth]{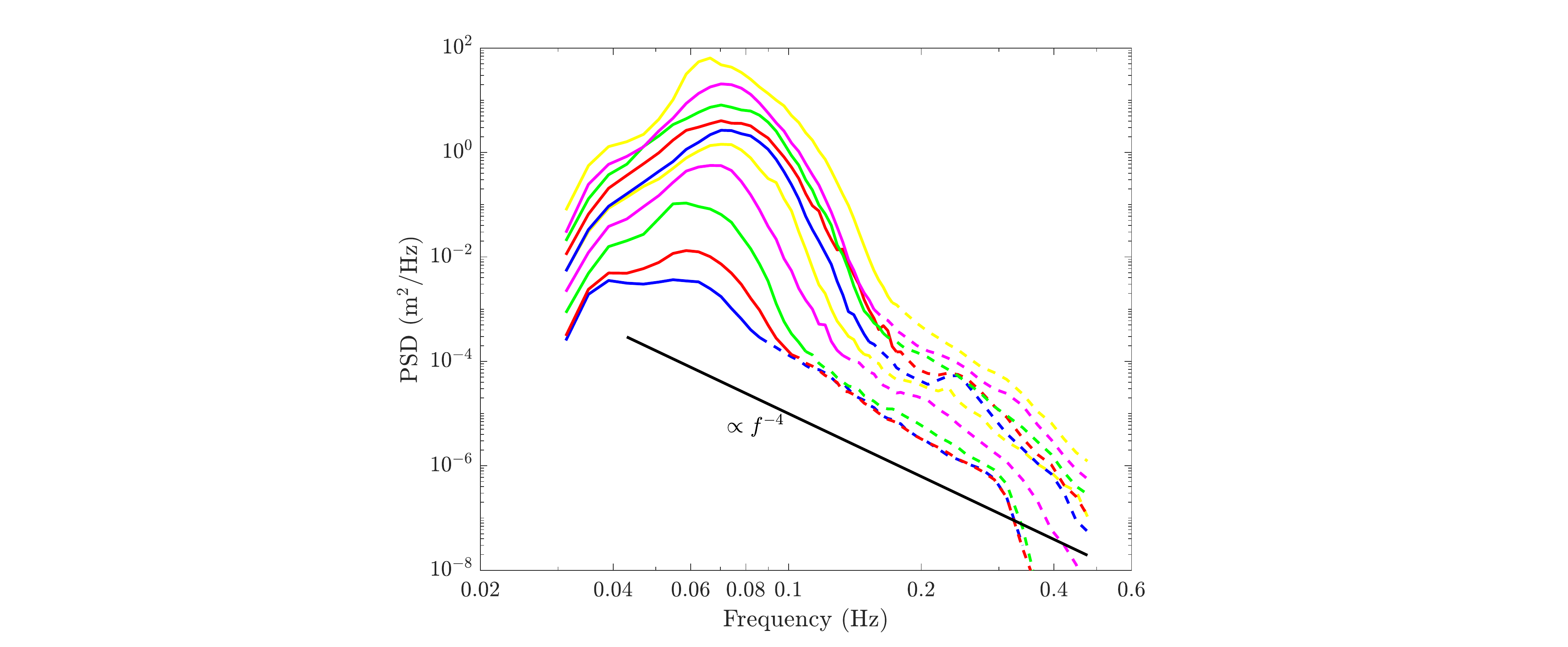}
    \caption{Average PSD in 10 groups of spectra recorded by a single buoy (part of the east deployment), and sorted according to SWH deciles (coloured curves). 
    The solid part of each curve corresponds to the non-filtered part of all spectra in the associated group, while the dashed part is filtered out.
    The figure uses a log-log scale.
    For reference, a curve with frequency dependence $f^{-4}$ is shown in solid black.}
    \label{fig2}
\end{figure}

Since our goal is to analyse the frequency-dependence of wave attenuation rates, we devised a method to identify $f_{cut}$ for each spectrum and filter out recorded spectral components for $f>f_{cut}$.
Our method is similar to that described by \cite{rogers_etal21}. 
For each buoy, the nine deciles of the SWH data are computed to bin the spectra into 10 groups with similar $H_s$ values.
The cutoff frequency $f_{cut}$ for all the spectra in each group is then identified manually by looking for a change of slope in the mean spectrum of that group.
Figure \ref{fig2} outlines the cutoff procedure for a single buoy that was part of the east deployment. 
The mean spectrum for each decile group is shown and $f_{cut}$ is chosen at the slope change points, such that for $f>f_{cut}$ (dotted lines) the spectral tail nearly follows the expected $f^{-4}$ frequency dependence caused by the instrument noise.  
These high-frequency tails are filtered out for all subsequent analyses.

\subsection{Attenuation coefficient}
\label{sec:an-attn}

Following standard wave data analysis techniques in the MIZ \citep[see, e.g.,][]{meylan_etal14}, we quantify the attenuation experienced by each component of the wave spectrum by making the following assumptions: (i) wave energy attenuates exponentially with distance travelled and (ii) for each 15-minute period, wave conditions are stationary. 
The rate of exponential wave energy attenuation at wave period $T=1/f$, referred to as the attenuation coefficient, is defined for each pair of buoys $i$ and $j$ as
\begin{equation}
    \label{eq:attn}
    \alpha(T) = \frac{\ln(S_i(T)) - \ln(S_j(T))}{D_{i,j}},
\end{equation}
where $S_i(T)$ and $S_j(T)$ are the PSD values of the two buoys in a given 15-minute period, such that buoy $i$ is located upstream in the direction of propagation of the wave at wave period $T$ and buoy $j$ is located downstream. 

The quantity $D_{i,j}$ in equation \eqref{eq:attn} is the effective separation between the two buoys measured along the incident wave direction. 
Since wave direction was not recorded by the buoys, we assume that all waves travel on an idealised north-to-south transect, which is consistent with most previous studies \citep[see, e.g.,][]{meylan_etal14}.
We will discuss how this choice of wave direction impacts attenuation estimates later in \textsection\ref{sec:dis}.\ref{sec:dis-sprd}.


During both deployments, the buoys drifted from their initial meridionally-spread configuration to a predominantly zonally-spread configuration, as discussed by \cite{kohout_etal20}.
As the angle between the buoys and the estimated incident wave direction, denoted by $\theta_{rel}$, increases from 0$^{\circ}$ (i.e.\ when the buoys are at the same longitude) up to $\pm 90^{\circ}$ (when the buoys are at the same latitude), the uncertainty on the estimated attenuation rate increases as $|\tan \theta_{rel}|$, which is singular in the latter configuration.
It is in fact straightforward to show that $\Delta\alpha=\alpha|\tan \theta_{rel}|\Delta\theta_{rel}$, where $\Delta\alpha$ and $\Delta\theta_{rel}$ are absolute uncertainty estimates.
Setting an exclusion criteria such that all attenuation coefficient measurements for which $|\theta_{rel}|<\theta_{max}$, for some prescribed $\theta_{max}$, are excluded from the analysis can mitigate this effect. 
However, in the context of the present deployments where the buoys are predominantly spread zonally, choosing $\theta_{max}$ to be too low will eliminate the vast majority of all attenuation coefficient measurements.
Here we set $\theta_{max}=75^{\circ}$. 
At this maximum angle, the relative uncertainty $\Delta\alpha/\alpha=100\%$ when the absolute uncertainty $\Delta\theta_{rel}=15^{\circ}$.


\subsection{Ice conditions}
\label{sec:an-conc}

The NOAA/NSIDC merged Climate Data Record of Passive Microwave Sea Ice Concentration \citep{peng_etal17,nsidc_conc_data} was used to estimate daily ice concentration changes on a polar stereographic grid with 25\,km spatial resolution. 
For each buoy pair and 15-minute observation period, sea ice concentration in the area of the buoys is estimated as the mean concentration in a domain bounded in latitudes by the location of the buoys and in longitudes by the location of the buoys plus half a degree in each direction.

We also use the in-situ ASPeCt data collected during each buoy deployment to estimate the ice conditions in the vicinity of each buoy.
To use these pointwise measurements in conjunction with wave attenuation data, we assume that the ice conditions observed at each deployment hold for a period of 24 hours post deployment.

\subsection{Wind conditions}
\label{sec:an-atm}

ERA5 reanalysis 10-m wind data \citep{era5_data} are analysed to understand the impact of local wind on observed wave attenuation rates in the MIZ.
The eastward and northward components of the wind velocity vector are given on a $0.25^{\circ} \times 0.25^{\circ}$ global grid.
We combine them into a 2-norm (Euclidean) wind speed metric and a wind direction metric (see \textsection\ref{sec:res}.\ref{sec:res-others}.\ref{sec:res-atm}).

\section{Results}\label{sec:res}

With all the filters described in \textsection\ref{sec:an}, we obtain 55398 and 190684 attenuation coefficient measurements for the west and east deployment, respectively.
Of all these, 6.8\% and 6.4\% are negative for the two deployments, respectively, corresponding to wave growth events.
Given the small proportion of observed wave growth events, we exclude them from all subsequent analyses.

\subsection{Attenuation as a function of period}\label{sec:res-freq}

The top panels in Figure \ref{fig1} show time-period spectrograms of the attenuation coefficient averaged over all available buoy pairs in successive 4-hour time windows.
The range of wave periods considered is $T=4\text{--}25$\,s.
For the west deployment (left panel), attenuation data were obtained for the first 8 days only.
After that time, two of the four buoys deployed failed, as discussed in \textsection\ref{sec:data}, and the two remaining buoys were too zonally spread to pass the exclusion criterion $\theta_{rel}<\theta_{max}$.
During these eight days, the average sea ice concentration did not exhibit significant changes, with $0.7\le c \le 0.8$. 
Similarly, the wave conditions remained fairly uniform and calm with an average measured SWH of 1\,m or less.
It is unfortunate that the significant freeze up event that took place after 27 April and the large wave event of 2 May were not included in the observational attenuation dataset, but relaxing further the exclusion criterion on $\theta_{rel}$ is not advisable given the fast increase of the uncertainty on the estimated attenuation rates as $\theta_{max}$ approaches $90^{\circ}$.

The observed attenuation coefficient consistently exhibits a peak at a wave period in the range 8--10\,s. 
Although this peak coincides with the high-frequency cutoff discussed in \textsection\ref{sec:an}.\ref{sec:an-filt} in some time intervals, in other intervals, attenuation decreases for short periods showing  that the spectral filtering may not have been sufficiently conservative to eliminate consistently the spurious high-frequency/short-period rollover effect.
In the second half of the 8-day record, we also observe an increase in the attenuation coefficient at long wave periods, which is not consistent with the expected decreasing trend of the attenuation coefficient with increasing wave period \citep{meylan_etal18}, noting that this effect has been observed previously by \cite{wadhams_etal88} who attributed it to the large noise-to-signal ratio in the short-period, low-energy tail of the recorded spectra. 


The east deployment (right panel) returned significantly more attenuation data, as more buoys survived longer.
The cutoff wave period seems to be somewhat correlated to ice concentration and SWH, which is expected given the relatively smaller signal-to-noise ratio of low-energy spectra which are more likely to be observed in highly concentrated sea ice. 
Up to 14 June, when the mean ice concentration was relatively high (i.e.\ $c>0.8$), the cutoff period was approximately 12\,s in calm conditions ($H_s<1$\,m) and 10\,s during the three wave events on 7, 9 and 11 June for which $H_s>1$\,m.
During each of these wave events, the attenuation coefficient seems to increase in all wave period bands, suggesting a nonlinear dependence of attenuation rates on wave energy.

In the following 12 days, the buoys drifted north and ice retreated causing the concentration to decrease ($c<0.8$) and the recorded waves to be of moderate energy with $H_s>1$\,m consistently during this period.
As a consequence, the cutoff wave period was at 5\,s, allowing us to analyse the short-period behaviour of ice-induced wave attenuation rates.
During that period, we observe peaks in attenuation at about 6\,s and again at about 13\,s, which is highly unusual and suggests different modes of attenuation were observed here, possibly depending on the wave and ice conditions.
Around 26 June, sea ice retreat in the sector of the east deployment (discussed in \textsection\ref{sec:data}), coinciding with an extremely large wave event ($H_s>7$\,m in the open ocean and $H_s>5$\,m in the MIZ), proved fatal for several buoys.
Only four buoys remained active after that event and were too zonally spread to return meaningful attenuation data. 
In the first week of July, however, two of the four buoys briefly realigned meridionally and provided more attenuation data as the sea ice expanded and wave activity was low.

\begin{figure}[t]
    \centering
    \includegraphics[width=1\textwidth]{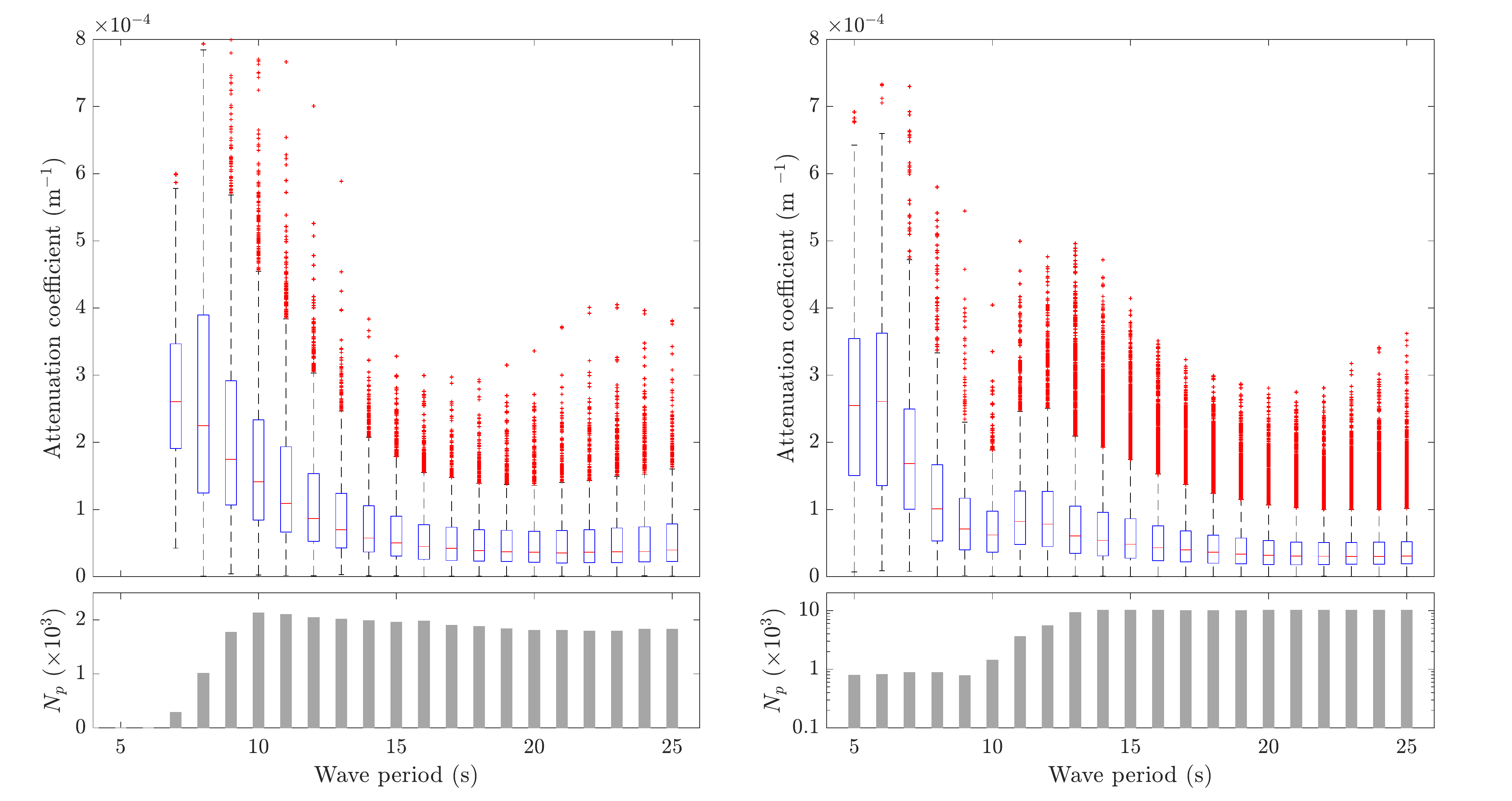}
    \caption{(Top panels) Box-and-whiskers plots of the attenuation coefficient as a function of wave period, binned into groups centered at integer values of the period in the range $5\le T\le 25$\,s.
    (Bottom panels) Number of attenuation coefficient estimates $N_p$ in each period bin. 
    Left and right pairs of panels correspond to west and east deployment data, respectively.
    }
    \label{fig3}
\end{figure}

Figure \ref{fig3} quantifies the dependence of the attenuation coefficient on wave period. 
All attenuation observations are binned into integer wave periods, i.e.\ $T= 4$, 5, 6, etc. 
For each wave period bin, the data are displayed in the form of a standard box-and-whisker plot (see top panels). 
The number of data points in each bin $N_p$ is also shown in the bottom panels.
For the west deployment (left panels), the median attenuation coefficient in each bin (red bar) decreases as wave period increases in a broad swell period range, i.e.\ $7\le T \le 21$\,s. 
Interestingly, no rollover is observed at short wave periods here, even though some evidence it could be present was seen in the spectrogram of figure \ref{fig1}.
For longer waves, i.e.\ $T>21$\,s, the attenuation coefficient seems to become nearly independent of wave period or slightly increases with it.

Overall, the dependence of the attenuation coefficient on wave period observed using west deployment data is fairly consistent with previous studies \citep[e.g.][]{meylan_etal14,rogers_etal21}.
In line with these studies, we fit an empirical power law to the median attenuation coefficient as a function of wave period.
This takes the form
\begin{equation}
    \alpha(T) = C T^{-\beta},
    \label{eq:power-law}
\end{equation}
where $C$ and $\beta$ are positive constants to be determined.
In the wave period range $7\le T \le 21$\,s, removing the influence of the nearly constant attenuation at long wave periods, standard least-square fitting gives $\beta\approx2.0$, which is on the lower end of values reported in the literature (note that the value of $C$ is not very informative and therefore not reported here).
Plotting $\alpha$ as a function of $T$ on a log-log scale (see figure \ref{fig5}, black dashed line on left panel), however, we observe clear deviations from the power law for short and long wave periods.
Restricting the power law fit to the wave period range $T=10$--$16$\,s, we obtain $\beta\approx2.5$, which is now consistent with most previous estimates \citep[see][]{meylan_etal18,montiel_etal18}.   

The right panels of figure \ref{fig3} display the attenuation data associated with the east deployment.
The two rollover peaks identified previously in the spectrogram (see figure \ref{fig1}, top-right panel), clearly emerge in the median estimates of the attenuation coefficients as a function of wave period, with a local maximum at $T=11\text{--}12$\,s and the global maximum at $T=6$\,s.
This pattern is not standard and difficult to explain, suggesting other physical quantities (e.g.\ ice conditions) may govern the $\alpha(T)$ functional dependence.
As a consequence, we do not attempt to fit a power law through the curve at this stage, as we will seek to untangle the different modes of attenuation underlying this observed pattern in \textsection\ref{sec:res}.\ref{sec:res-others}. 
We note that the number of observations at short wave periods, i.e.\ $T\le10$\,s, is significantly smaller than for longer periods, corresponding to a wider spread of attenuation rate estimates and therefore a larger uncertainty.
The long-period tail (i.e.\ $T>20$\,s) of the median attenuation coefficient estimates also tends to plateau, which is consistent with our observations on west deployment data.
Superimposing the median curves obtained for the two deployments (not shown here) actually shows that the long-period behaviour is remarkably similar in magnitude as well, with the median attenuation rates obtained for the two deployments nearly coinciding in the regime $T\ge12$\,s.

\subsection{Effect of ice, wave and wind conditions}\label{sec:res-others}

\subsubsection{Ice concentration}\label{sec:res-conc}

\begin{figure}[t]
    \centering
    \includegraphics[width=1\textwidth]{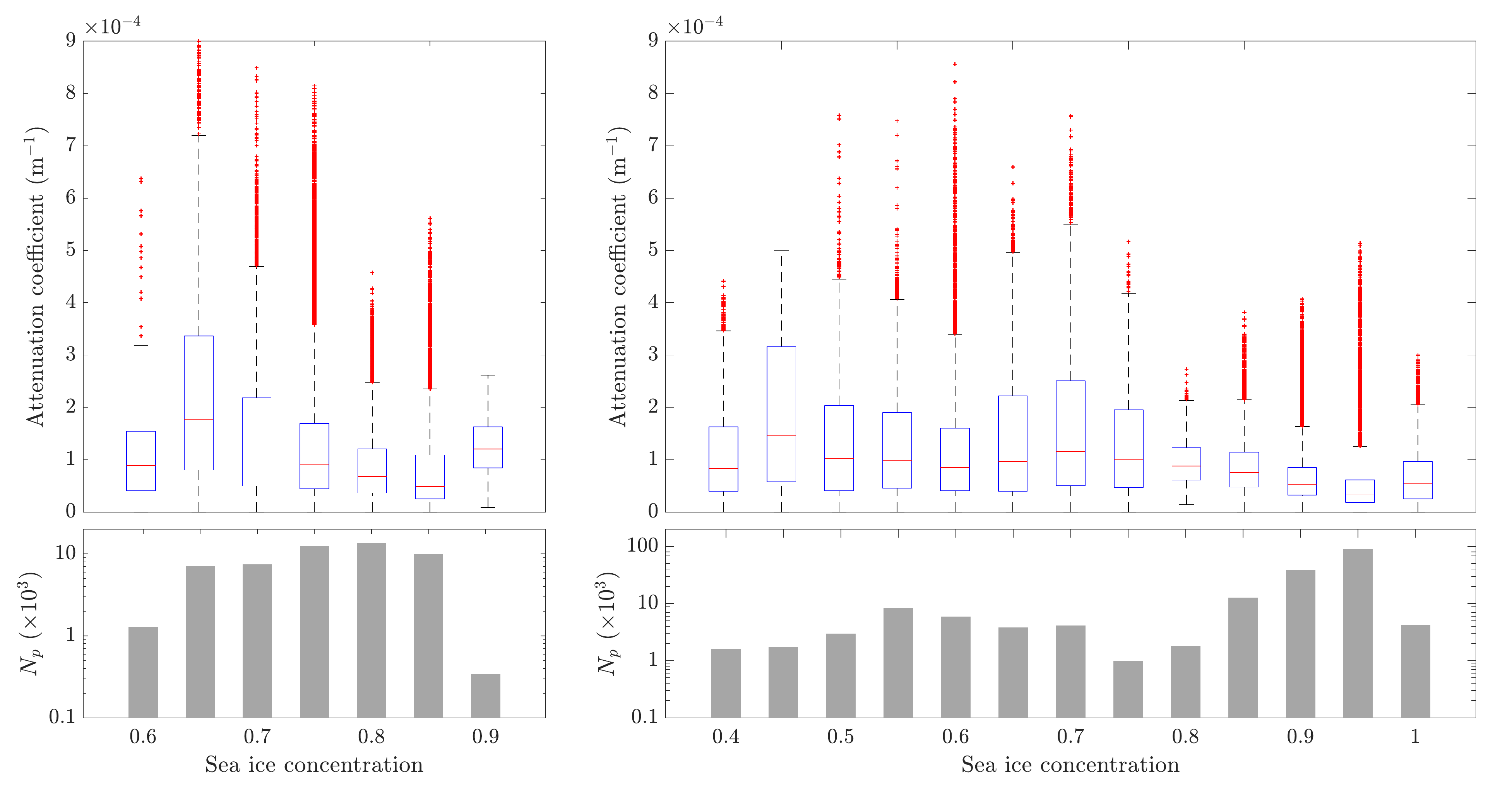}
    \caption{(Top panels) Box-and-whiskers plots of the attenuation coefficient as a function of sea ice concentration, binned into groups centered at integer multiples of 0.05 in the range $0\le c\le 1$.
    (Bottom panels) Number of attenuation coefficient estimates $N_p$ in each ice concentration bin. 
    Left and right pairs of panels correspond to west and east deployment data, respectively.}
    \label{fig4}
\end{figure}

Figure \ref{fig4} shows the distribution of measured attenuation coefficients in ice concentration bins of width 0.05 (measured using the method discussed in \textsection\ref{sec:an}.\ref{sec:an-conc}) for both the west and east deployments (left and right panels, respectively).
During the west deployment, attenuation data were obtained for a restricted range of concentrations ($0.6\le c\le 0.9$), with the most common data ($\sim 69\%$) in the range $0.75\le c\le0.85$.
During the east deployment, a much broader range of ice concentrations were encountered ($0.4\le c\le 1$), although more than half were captured for ice concentration $c\approx0.95$ and three quarters for $0.9\le c\le1$. 
This dominant ice condition can be explained by the fact that the buoys have been deployed relatively far from the ice edge in a closed pack.
As the buoys later drifted north towards the ice edge, a few of them failed and the others shifted to a predominantly zonally-spread configuration (as discussed in \textsection\ref{sec:an}.\ref{sec:an-attn}), for which attenuation estimates were excluded.

The most striking feature in figure \ref{fig4} is the lack of a simple pattern describing the effect of sea ice concentration on wave attenuation rates.
More specifically, the attenuation coefficient does not seem to increase with ice concentration. 
If anything, the data show an inverse relationship, although it is not a significant effect.
We also looked at how attenuation rates depend on the mean distance of the buoys to the ice edge, $d_{\mathrm{edge}}$, as that quantity is strongly correlated to ice concentration.
The decreasing trend of $\alpha$ with respect to $d_{\mathrm{edge}}$ is confirmed (see supplementary figure S1).

\begin{figure}[t]
    \centering
    \includegraphics[width=1\textwidth]{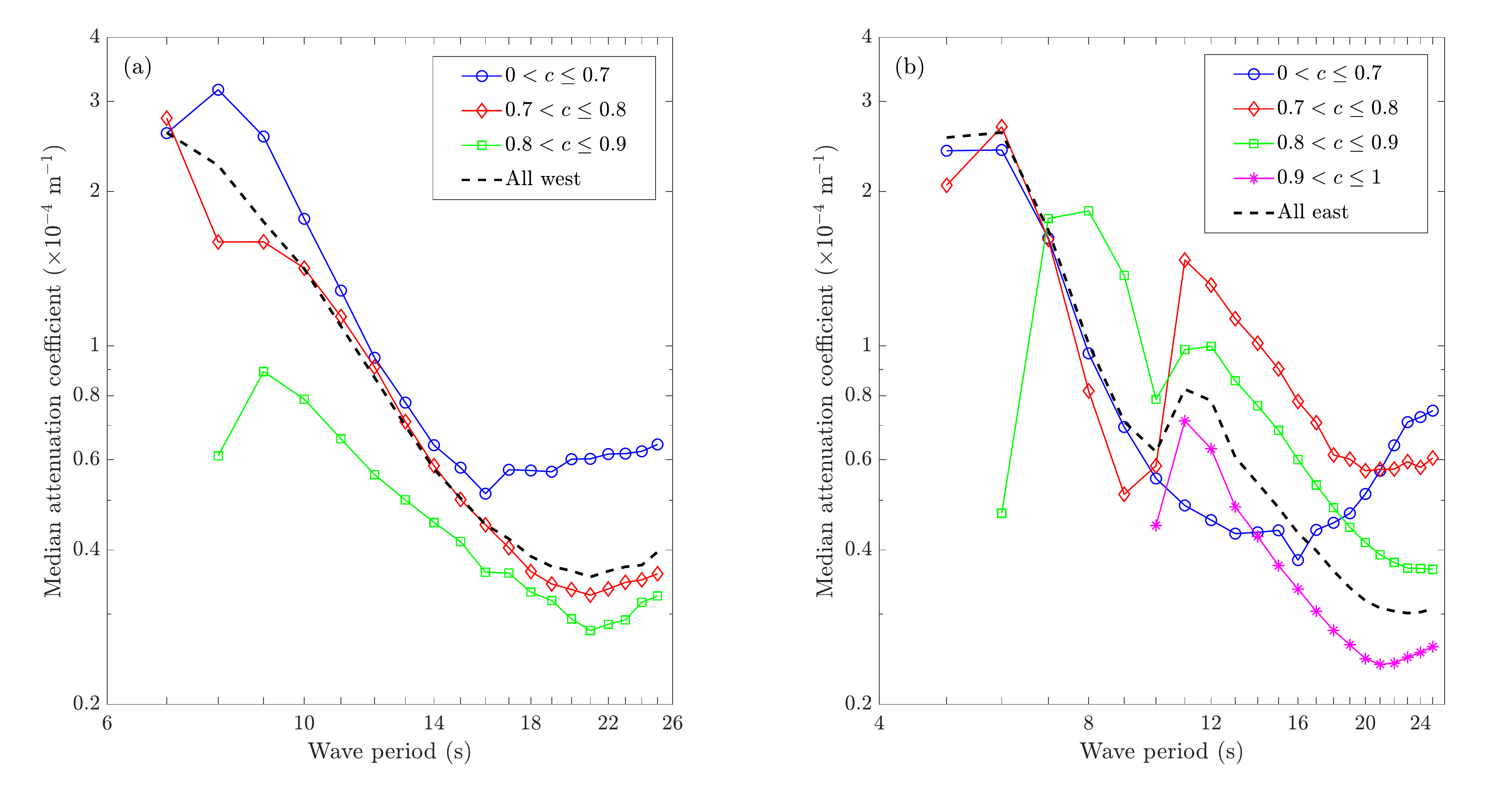}
    \caption{Median value of the attenuation coefficient binned by ice concentration and wave period, as a function of wave period, and plotted on a log-log scale (coloured curves with markers).
    The corresponding curve obtained when combining all concentration bins is shown as a dashed black line.
    Panels (a) and (b) correspond to west and east deployment data, respectively.
    }
    \label{fig5}
\end{figure}

To understand better how ice concentration affects attenuation in different parts of the wave spectrum, we further bin the attenuation data in each concentration bin into the wave period groups considered in \textsection\ref{sec:res}.\ref{sec:res-freq}.
Figure \ref{fig5} shows the median attenuation coefficient as a function of wave period for different ice concentration regimes.
A log-log scale is used here to show the power law dependence discussed previously.
For the west deployment (left panel), the decrease in attenuation rates with ice concentration can be observed across the entire range of wave periods.
At low concentration (i.e.\ $c\le0.7$), we observe a rollover peak at $T=8$\,s, above which the attenuation coefficient seems to depend on period via a power law, up to $T=16$\,s.
For longer wave periods, the attenuation plateaus or even increases slightly.
The same qualitative behaviour is seen for the other concentration bins, although the plateauing seems to shift to longer wave periods for higher ice concentrations.
We will later argue that this observed feature is likely to be an artefact due to the limitations of our approach to estimate attenuation rates at long wave periods.
We can, however, fit a power law of the form given by eq.~\eqref{eq:power-law} for each ice concentration band in a limited range of wave periods.
The results of this procedure are summarised in table \ref{tab1}.
The power parameter $\beta$ seems to decrease with concentration.

For the east deployment, two different regimes of attenuation seem to emerge, i.e.\ for short and long wave periods. 
Attenuation rates for short periods, i.e.\ up to approximately 10\,s, were measured almost exclusively at low ice concentration (i.e.\ $c\le0.8$). 
A power-law fit in the period range $6\le T\le 10$\,s for $c\le0.8$ gives $\beta\approx3.0$.  
This regime captures attenuation close to the ice edge, where relatively low-concentration sea ice quickly dampens the short-period components of the spectrum due to scattering and other dissipation mechanisms \citep{squire_montiel16}.
Deeper into the MIZ, these components are essentially removed from the recorded spectra, so that attenuation of long-period components (i.e.\ $T\ge12$\,s) only can be measured. 
Similarly to the west deployment, we observe an inverse relationship between attenuation rates and ice concentration in this regime.
Power law fits in each concentration band give very similar estimates for the parameter $\beta$ (see table \ref{tab1}), suggesting that the same dissipation processes cause the observed damping. 

\begin{table}[h]
\begin{center}
\begin{tabular}{c||c|c|c||c|c|c|c|}
   \multicolumn{1}{c}{}  & \multicolumn{3}{c}{West deployment} & \multicolumn{4}{c}{East deployment} \\
                     & $0< c \le 0.7$ & $0.7< c \le 0.8$ & $0.8< c \le 0.9$ & $0< c \le 0.7$ & $0.7< c \le 0.8$ & $0.8< c \le 0.9$ & $0.9< c \le 1$ \\
     \hline
    $\beta$          & $2.8$    & $2.3$    & $1.4$    & $3.0$ & $1.8$ & $1.7$ & $1.8$ \\
    Period range (s) & $[8,16]$ & $[9,19]$ & $[9,21]$ & $[6,10]$ & $[12,18]$ & $[12,20]$ & $[12,20]$ 
\end{tabular}
\end{center}
\label{tab1}
\caption{Power-law parameter $\beta$ fitted to $\alpha(T)$ curves for each deployment and ice concentration bin.
The corresponding range of wave periods on which the fit was achieved is also indicated in each case.}
\end{table}

\subsubsection{Significant wave height and mean period}\label{sec:res-ww3}

\begin{figure}[t]
    \centering
    \includegraphics[width=1\textwidth]{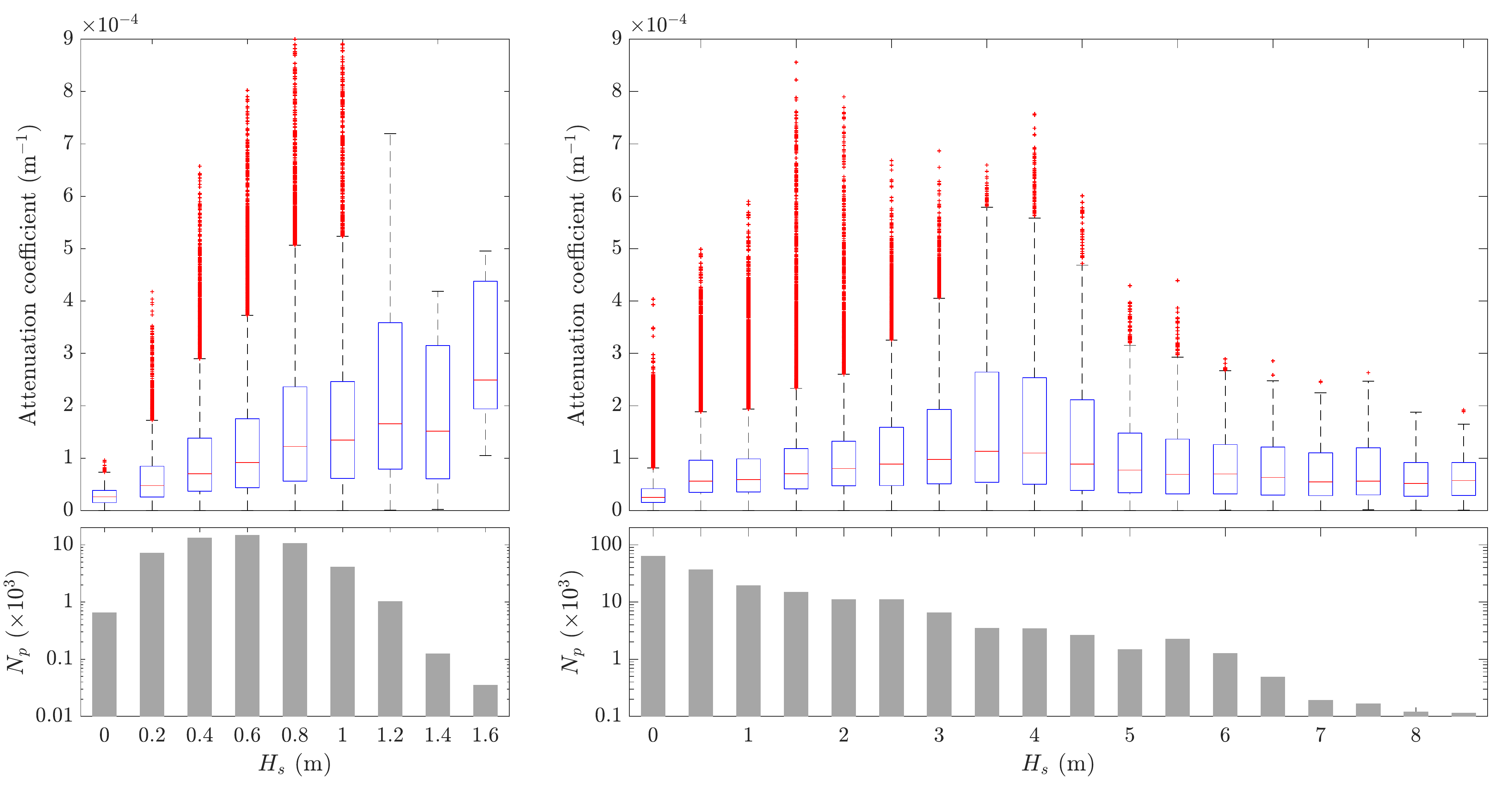}
    \caption{Same as figure~\ref{fig4}, but with attenuation coefficients binned by SWH $H_s$.}
    \label{fig6}
\end{figure}

\begin{figure}[t]
    \centering
    \includegraphics[width=1\textwidth]{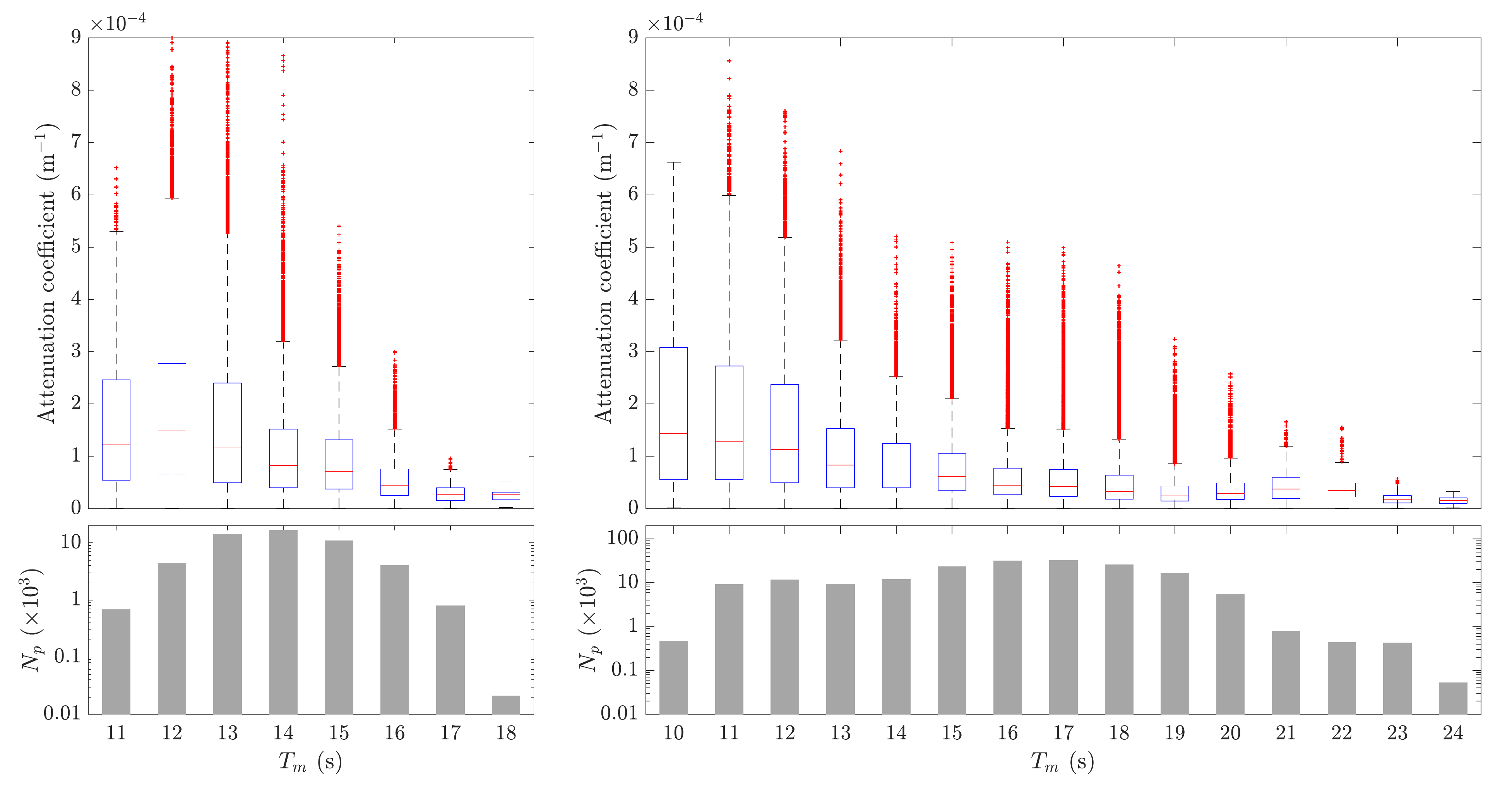}
    \caption{Same as figure~\ref{fig4}, but with attenuation coefficients binned by mean wave period $T_m$.}
    \label{fig7}
\end{figure}

Figures \ref{fig6} and \ref{fig7} show attenuation rate data binned by SWH and mean wave period, respectively.
For each attenuation rate estimate, the SWH and mean period are computed as the arithmetic mean of those quantities for the two buoys involved.
The mean wave period for each buoy $i$, in each 15-minute observation window, is defined as 
\[
T_m = \frac{\int_{f_{min}}^{f_{max}} S_i(f)\,df}{\int_{f_{min}}^{f_{max}} S_i(f)f\,df}.
\]
As discussed previously, very calm wave conditions were encountered during the west deployment with SWH well below 2\,m for most of the buoy pairs used to compute attenuation rates (see figure \ref{fig6}, left panels).
The median attenuation rate increases nearly monotonically with SWH in this range of SWHs, which is sensible given that energetic wave events are more likely to contain short-period spectral components, which attenuate faster, than calmer events.
This claim is further reinforced by inspecting the dependence of attenuation rates on the mean wave period (see figure \ref{fig7}, left panels), which shows attenuation rates monotonically decreasing as the mean period increases.

The broader range of wave conditions observed during the east deployment (see figures \ref{fig6} and \ref{fig7}, right panels) show that attenuation rates do increase with SWH
up to 3--4\,m, but then monotonically decrease for larger wave events, noting that significantly less data were recorded in this latter regime.
The small-wave behaviour (i.e.\, $H_s<3$\,m say) can be explained by the relatively lower short-period spectral content of smaller waves, which have already been filtered out through ice-induced attenuation.
For large wave events (i.e.\, $H_s>4$\,m say) which have not been significantly affected by the presence of sea ice, e.g.\ close to the ice edge, the SWH wave height increases with the mean period (actually $H_s \propto T_m^2$ for a Pierson-Moskowitz spectrum), so that the short-period content of larger wave events is relatively lower than that of smaller wave events. 
Given these two extreme behaviours, the observed peak in attenuation rates for $H_s\approx3$\,m can be expected.
The near-monotonic decrease of attenuation rates with mean wave period (see figure \ref{fig7}) confirms that the magnitude of ice-induced wave attenuation is primarily governed by the short-period spectral content of those waves.

\begin{figure}[t]
    \centering
    \includegraphics[width=1\textwidth]{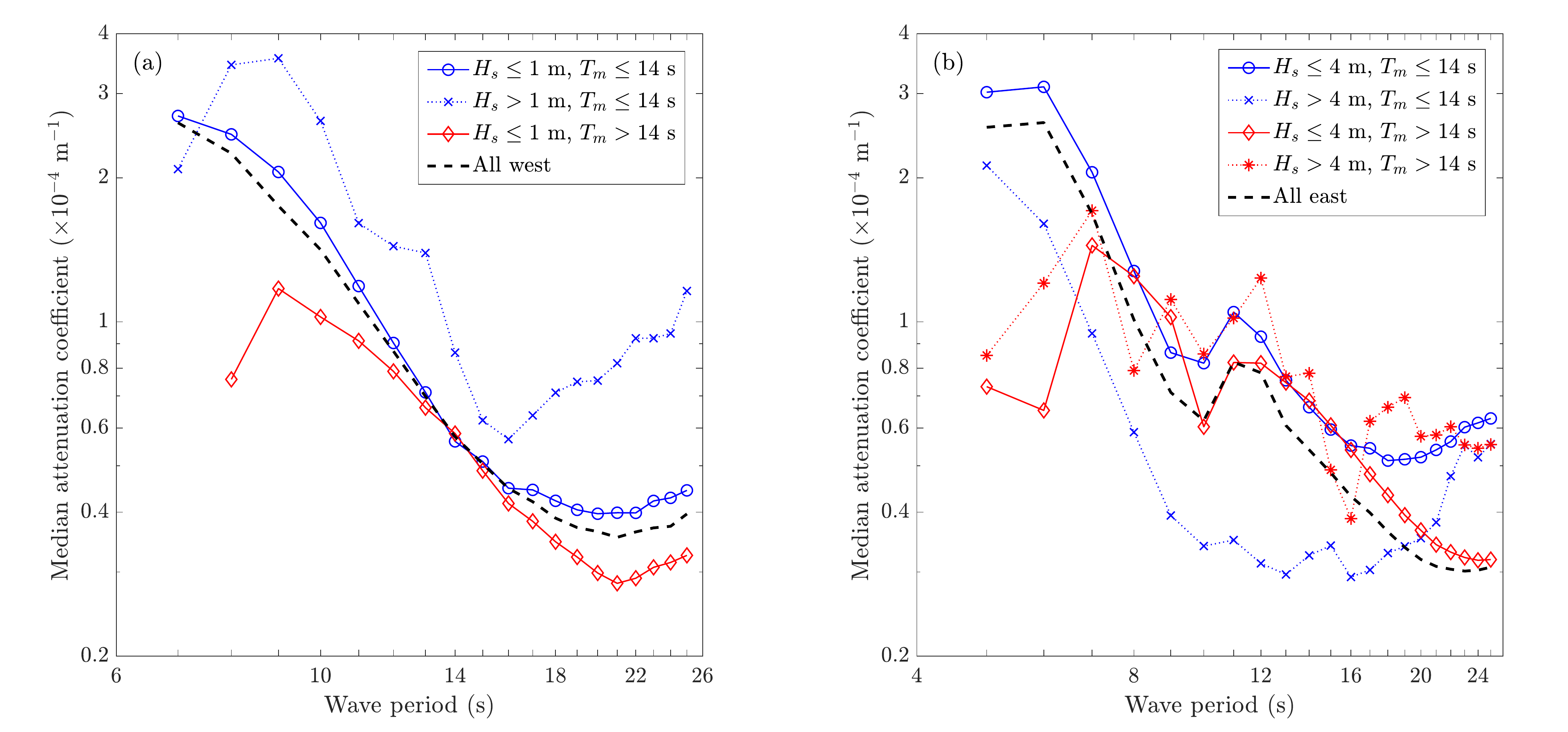}
    \caption{Same as figure~\ref{fig5}, but with attenuation coefficients binned by SWH $H_s$ and mean wave period $T_m$ (4 groups), instead of concentration.}
    \label{fig8}
\end{figure}

In figure \ref{fig8}, the median attenuation coefficient is plotted as a function of wave period for four different wave regimes.
The results of the power-law fit for each case are summarised in table \ref{tab2}.
The left panel considers the west deployment, for which we look at the influence of SWH and mean wave period combined.
The majority of the attenuation data were obtained for $H_s\le1\,$ and $T_m\le14$\,s.
In this case, the $\alpha(T)$ curve (solid blue line with circle markers) nearly matches that obtained when considering all attenuation data measured during the west deployment.
A power-law fit in the period range $T=9$--16\,s gives $\beta\approx2.8$, which is slightly higher than that obtained with all the data.
For larger waves ($H_s>1$\,m), while keeping $T_m\le14$\,s, the magnitude of the attenuation coefficient increases across all wave periods, noting a strong rollover for $T<9$\,s and again for $T>16$\,s.
In the $T=9$--16\,s range, however, the power-law fit yields $\beta\approx3.2$, which is similar to that obtained for small waves.
This suggests that SWH causes the $\alpha(T)$ to shift upward, without significant changes to its slope.
Now looking at small waves ($H_s<1$\,s) with larger mean wave period ($T_m>14$\,s), we observe a downward shift of the $\alpha(T)$ curve, along with a significant decrease of the slope in the power-law relationship ($\beta<2$).
Note that no attenuation data were obtained when $H_s>1\,$ and $T_m>14$\,s.

The right panel of figure \ref{fig8} shows median $\alpha(T)$ curves obtained when the east deployment attenuation data are grouped according to their location in the $(H_s,T_m)$-plane.
For $T_m\le14$\,s, we observe a strong influence of the SWH with attenuation rates consistently higher in magnitude for smaller wave events (i.e.\ $H_s\le 4$\,m) than for larger wave events across the range of wave periods.
Power-law fits for $T=6$--10\,s give $\beta\approx2.8$ and 3.1, respectively (see table \ref{tab2}), which are similar values to those found using west deployment attenuation data for $T_m\le14$\,s.
We note, however, that the range of wave periods used for the fit is different.
For $T_m>14$\,s, we observe much less influence of the SWH on the attenuation rates.
It should be noted that over 80\% of the east deployment attenuation data were recorded in this regime for $H_s\le 4$\,m.
In this case, we see evidence of the spurious rollover effect for $T<12$\,s and $T>24$\,s.
A power-law fit for $T=13$--20\,s gives $\beta\approx1.7$ (see table \ref{tab2}), which is similar to the value obtained in this regime using west deployment data. 
The data obtained for the large wave events ($T_m>14$\,s and $H_s>4$\,m) only represent $0.25\%$ of all east deployment attenuation rate estimates. 
Although the corresponding $\alpha(T)$ in figure \ref{fig8} follows a similar pattern as that obtained for $H_s\le4$\,m, the plateauing $\alpha$ values for $T>16$\,s does not allow us to fit a meaningful power-law through the data.

\begin{table}[h]
\label{tab2}
\begin{center}
\begin{tabular}{c||c|c|c||c|c|c|c|}
   \multicolumn{1}{c}{}  & \multicolumn{3}{c}{West deployment} & \multicolumn{4}{c}{East deployment} \\
                     & $H_s\le1$\,m & $H_s>1$\,m & $H_s\le1$\,m & $H_s\le4$\,m & $H_s>4$\,m & $H_s\le4$\,m & $H_s>4$\,m \\
                     & $T_m\le14$\,s & $T_m\le14$\,s & $T_m>14$\,s & $T_m\le14$\,s & $T_m\le14$\,s & $T_m>14$\,s & $T_m>14$\,s \\
     \hline
    $\beta$          & $2.8$    & $3.2$    & $1.8$    & $2.8$ & $3.1$ & $1.7$ & X \\
    Period range (s) & $[9,16]$ & $[9,16]$ & $[9,21]$ & $[6,10]$ & $[6,10]$ & $[13,20]$ & X
    
\end{tabular}
\end{center}
\caption{Same as table~\ref{tab1} but for attenuation data grouped into SWH/mean wave period bins.
}
\end{table}

\subsubsection{Wind speed and direction}\label{sec:res-atm}

\begin{figure}[t]
    \centering
    \includegraphics[width=1\textwidth]{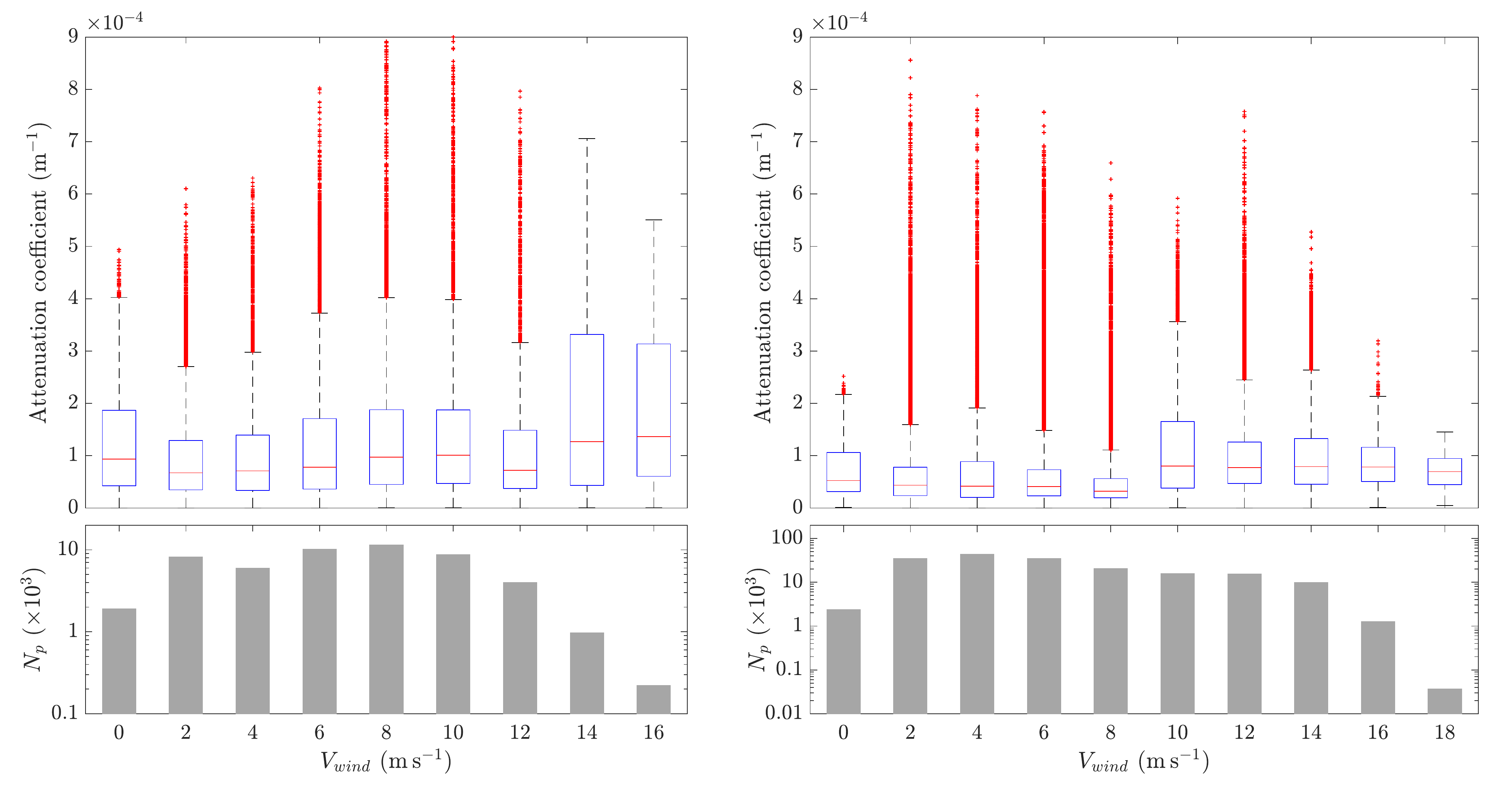}
    \caption{Same as figure~\ref{fig4}, but with attenuation coefficients binned by wind speed $V_{wind}$.}
    \label{fig9}
\end{figure}

\begin{figure}[t]
    \centering
    \includegraphics[width=1\textwidth]{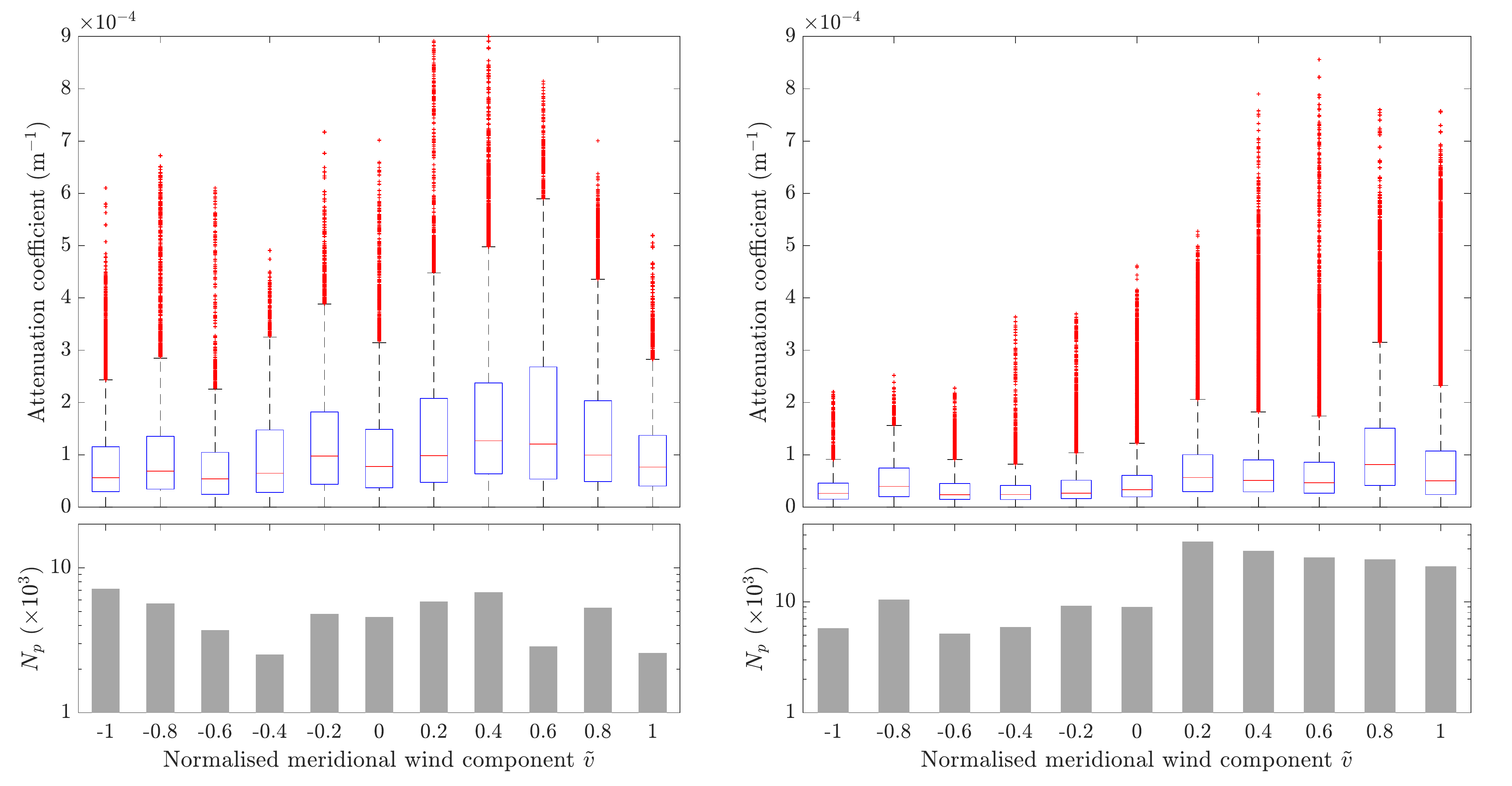}
    \caption{Same as figure~\ref{fig4}, but with attenuation coefficients binned by normalised meridional wind component $\tilde{v}$.}
    \label{fig10}
\end{figure}

We now turn our attention to the effect of wind speed and direction on observed attenuation rates. 
Given the wind velocity vector $(u,v)$, where $u$ and $v$ are the zonal and meridional components, respectively, we define wind speed $V_{wind}$ as the Euclidean 2-norm of the vector, i.e.\ $V_{wind} = \left(u^2 + v^2\right)^{1/2}$. 
Wind direction is quantified by the normalised meridional component of the wind vector, i.e.\ $\tilde{v} = v/V_{wind}\in[-1,1]$, where $\tilde{v}=-1$ and $\tilde{v}=1$ correspond to northerly and southerly winds, respectively.
Figures \ref{fig9} and \ref{fig10} show the attenuation rate estimates binned by $V_{wind}$ and $\tilde{v}$, respectively.
Again, the arithmetic mean of these quantities for the two buoys involved is used here.
For both deployments, we observe a general increase of the attenuation rates with increasing wind speed.
West deployment data indicate a gradual increase, while east deployment data seem to show an abrupt increase at $V_{wind}\approx10$\,m\,s$^{-1}$, so that attenuation rates are nearly constant for smaller and larger wind velocities.
Note that the number of attenuation rate estimates for high wind speeds is much smaller than for low and mid-range wind speeds.

Some effect of wind direction can also be observed on attenuation rate data (see figure \ref{fig10}).
Specifically, we can see that attenuation rates are positively correlated with the normalised meridional wind component $\tilde{v}$.
This is consistent for both deployments.
Although wind-induced wave generation in ice-covered sea is poorly understood, this effect intuitively makes sense, as northerly-dominant winds ($\tilde{v}<0$) should reinforce south-travelling waves propagating in the same direction.
Conversely, southerly-dominant winds ($\tilde{v}>0$) are expected to induce a resistive stress on south-travelling waves, therefore causing an apparent increase in wave attenuation.
Although during the west deployment no specific wind condition seem to dominate, southerly-dominant winds were clearly more prevalent during the east deployment (see lower panels in figure \ref{fig10}).

\begin{figure}[t]
    \centering
    \includegraphics[width=1\textwidth]{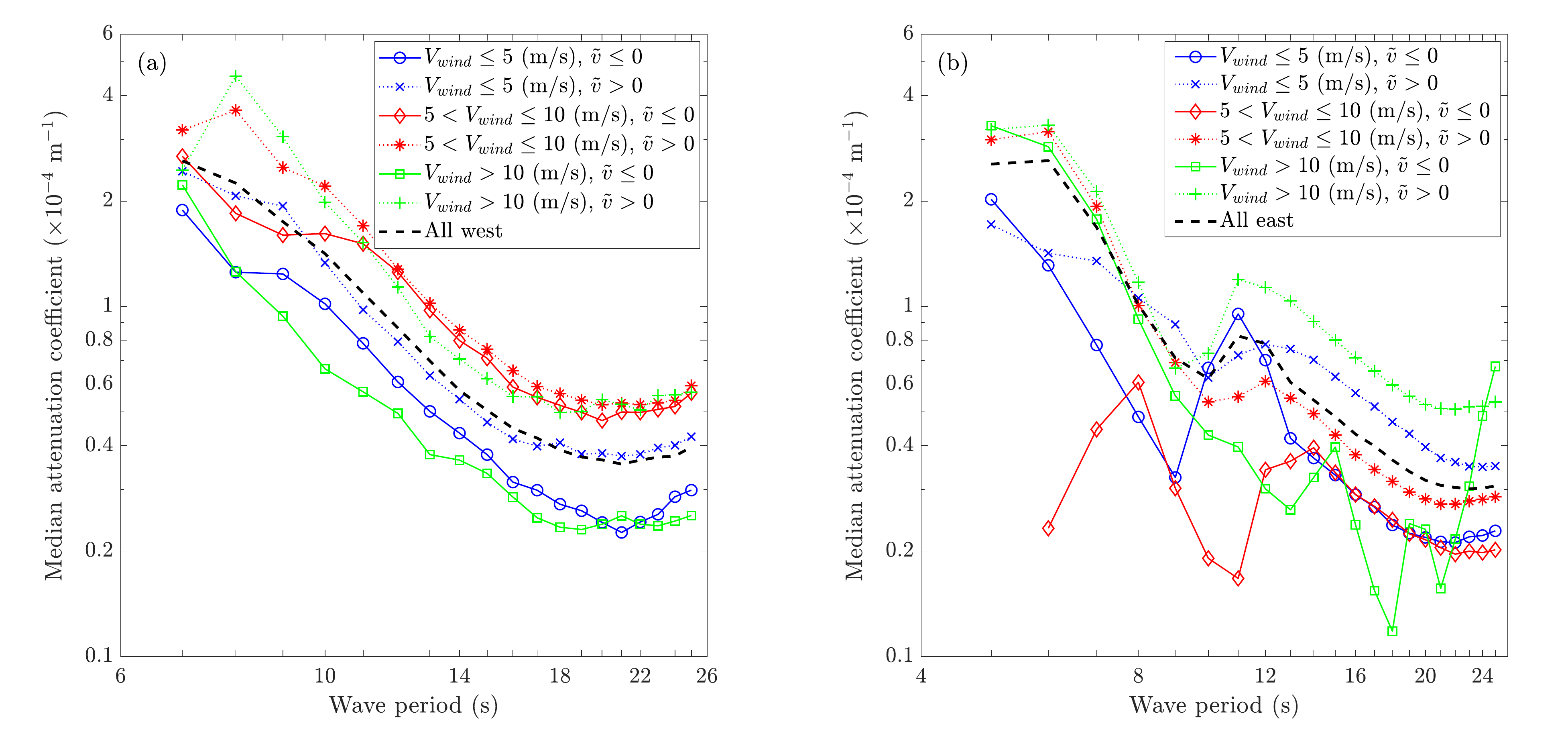}
    \caption{Same as figure~\ref{fig5}, but with attenuation coefficients binned by wind speed $V_{wind}$ and normalised meridional wind component $\tilde{v}$ (6 groups), instead of concentration.}
    \label{fig11}
\end{figure}

Figure \ref{fig11} shows median attenuation rates as a function of wave period for different wind regimes and both deployments.
Six wind regimes are considered here, i.e.\ calm wind ($V_{wind}\le5$\,m\,s$^{-1}$), intermediate wind ($5<V_{wind}\le10$\,m\,s$^{-1}$) and strong wind ($V_{wind}>10$\,m\,s$^{-1}$), for both northerly-dominant and southerly-dominant wind directions.
For any given wind-speed regime, we observe consistently larger attenuation rates for southerly-dominant wind conditions. Interestingly, this magnification does not seem to affect any preferred band of wave periods.
The influence of wind speed on attenuation rates, if any, is harder to decipher.
Stronger winds do not seem to be consistently associated with larger attenuation rates and do not affect specific wave period bands.
As a consequence, there is no real advantage in fitting power-laws to the different $\alpha(T)$ curves here. 
 
\section{Discussion}\label{sec:dis}

\subsection{Drivers of $\alpha(T)$}\label{sec:dis-proc}

In the previous section, we have identified three important physical drivers of ocean wave attenuation in the MIZ: (i) ice concentration, (ii) mean wave period and (iii) wind direction.
Specifically, the attenuation rate $\alpha$ seems to be inversely correlated to concentration, which also affects the power-law relationship $\alpha \propto T^{-\beta}$ with $\beta\sim3$ at low concentration and $\beta<2$ at high concentration.
We also observed an inverse correlation between the attenuation rate and mean wave period, such that $\beta\sim3$ for $T_m<14$\,s and $\beta<2$ for $T_m>14$\,s.
Finally, wind direction significantly affects the magnitude of wave attenuation, with northerly-dominant and southerly-dominant winds reducing and increasing attenuation, respectively, but without significant effects on its power-law relationship to wave period.
Our analysis suggests that ice, wave and wind conditions all play a significant role in explaining observed wave attenuation rates in the MIZ.

\begin{figure}[t]
    \centering
    \includegraphics[width=1\textwidth]{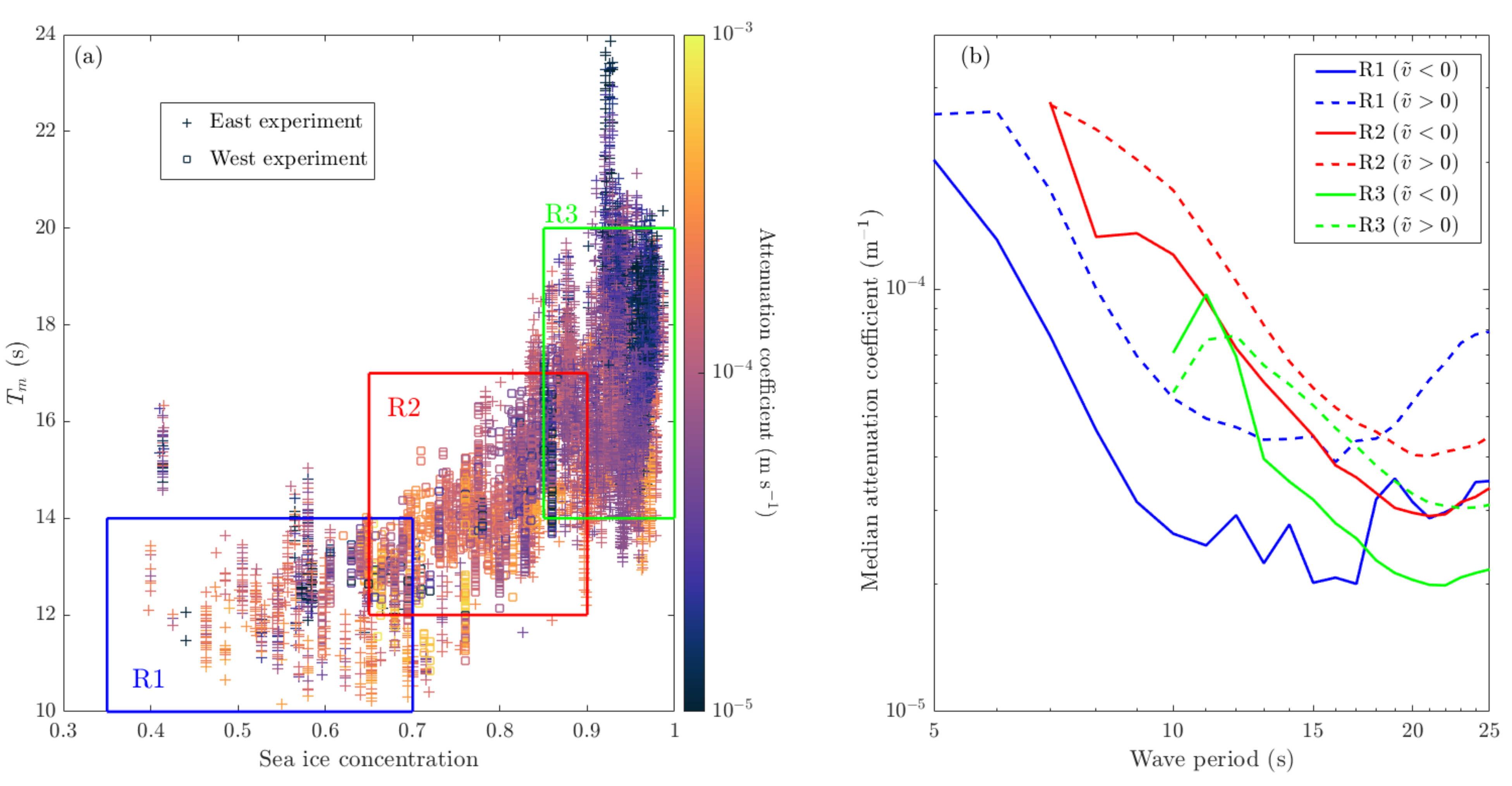}
    \caption{(a) Scatter plot of all attenuation coefficient estimates (circle and plus markers correspond to west and east deployment data, respectively) in the $c$-$T_m$ plane. 
    The colour of each marker describes its magnitude.
    The three rectangular regions labelled R1, R2 and R3, identify the three regimes of ice-induced wave attenuation discussed in the text.
    (b) Median attenuation coefficient in each wave period bin, as a function of wave period, for attenuation data in the three regimes (blue, red and green for R1, R2 and R3, respectively) and further grouped into wind direction bins (dashed and solid curves correspond to southerly-dominant, $\tilde{v}>0$, and northerly-dominant, $\tilde{v}<0$, wind conditions, respectively).}
    \label{fig12}
\end{figure}

To understand these relationships in more depth, we look at the correlation between sea ice concentration and mean wave period. 
Figure \ref{fig12}(a) shows the corresponding scatter plot for all attenuation coefficient estimates, for both the west and east deployments (squares and pluses, respectively).
We observe a clear positive correlation between sea ice concentration and mean wave period overall.
Careful inspection of the plot reveals that east deployment data are clustered into two clearly separated groups in the $(c,T_m)$ plane.
The first group, referred to as regime 1 (R1), is located on the lower-left region of the plane, and can be approximately defined by $0.35\le c\le0.7$ and $10\le T_m\le14$\,s (see blue box).
The second group, referred to as regime 3 (R3), is located on the mid/upper-right region of the plane.
Excluding extreme $T_m$ values, we define this regime by $0.85\le c\le 1$ and $14< T_m\le20$\,s (see green box).
Interestingly, these two disjoint clusters are bridged by the west deployment data, which we use to define regime 2 (R2), such that $0.65\le c\le 0.9$ and $12< T_m\le17$\,s (see red box).
Note that although R2 intersects both R1 and R3, only west deployment data are used to define it, while only east deployment data are used to define the two latter regimes, so no attenuation measurements belong to more than one regime.
We also note that this clustering accounts for approximately 88\% of the east deployment data and 86\% of the west deployment data.

The marker color in figure \ref{fig12}(a) quantifies the attenuation coefficient of each measurement. 
Overall, larger attenuation coefficient values seem to be observed in R1 and R2 than in R3, noting that the sample size is much smaller in these two former regimes than in the latter one.
To decipher the regime-dependence of attenuation coefficient estimates, in figure \ref{fig12}(b) we plot the median attenuation rate as a function of wave period for all three regimes, further binning the data into northerly-dominant ($\tilde{v}\le0$) and southerly-dominant ($\tilde{v}>0$) wind directions.
We observe that the short-period rollover nearly disappears, as it can only be observed in R1 (with $\tilde{v}>0$) for $T<6$\,s, and in R3 for $T<13$\,s. 
For long wave periods, however, we observe a clear trend reversal and deviation from the power law, which is likely caused by the low signal-to-noise ratio in the underlying spectra. 
As in \textsection\ref{sec:res}, we fit power-law relationships of the form \eqref{eq:power-law} to each curve in a limited range of $T$ values. 
The results of the fit are summarised in Table \ref{tab3}.
For each regime, we also include the mean and standard deviations of the SWH $H_s$, and the distance to the ice edge $d_{\mathrm{edge}}$.
Full distributions of these quantities and a few others are shown in the supplementary figures S3--S5.

\begin{table}[h]
\label{tab3}
\begin{center}
\begin{tabular}{c||c|c||c|c||c|c|}
    \multicolumn{1}{c}{}             & \multicolumn{2}{c}{R1}          & \multicolumn{2}{c}{R2}          & \multicolumn{2}{c}{R3} \\
                                     & $\tilde{v}\le0$ & $\tilde{v}>0$ & $\tilde{v}\le0$ & $\tilde{v}>0$ & $\tilde{v}\le0$ & $\tilde{v}>0$ \\     
     \hline
     $\beta$                         & $2.9$           & $2.9$         & $2.0$           & $2.0$         & $1.5$           & $1.5$ \\
     $\ln C$                         & $-3.9$          & $-3.1$        & $-4.4$          & $-4.1$        & $-6.3$          & $-5.7$ \\
     $T$ range                       & $[5,11]$        & $[6,11]$      & $[7,20]$        & $[7,20]$      & $[13,21]$       & $[12,23]$ \\
     Number of attenuation estimates & $538$           & $9951$        & $13461$         & $12722$       & $32286$         & $80875$ \\
     \hline
     $H_s$ (mean; in m)                    & \multicolumn{2}{c||}{5.5}         & \multicolumn{2}{c||}{0.5}         & \multicolumn{2}{c|}{0.6} \\
     $H_s$ (STD; in m)                     & \multicolumn{2}{c||}{1.4}         & \multicolumn{2}{c||}{0.2}         & \multicolumn{2}{c|}{0.7} \\
     $d_{\mathrm{edge}}$ (mean; in km)                    & \multicolumn{2}{c||}{40}         & \multicolumn{2}{c||}{65}         & \multicolumn{2}{c|}{129} \\
     $d_{\mathrm{edge}}$ (STD; in km)                     & \multicolumn{2}{c||}{14}         & \multicolumn{2}{c||}{21}         & \multicolumn{2}{c|}{52}
\end{tabular}
\end{center}
\caption{Power-law fit parameters $\beta$ and $C$ on the $\alpha(T)$ curves shown in figure~\ref{fig12}(b) for the three identified regimes of wave attenuation and both northerly-dominant and southerly-dominant wind conditions.
For each regime, the mean and standard deviation (STD) of the significant wave height and distance to the ice edge are indicated as references.}
\end{table}

We observe a clear effect of the ice and wave conditions on the power-law fit, as the slope (governed by $\beta$) clearly decreases as the ice concentration and mean period increase (i.e.\ going from R1 to R3).
The wind direction, on the other hand, does not affect the slope but causes a vertical shift of the $\alpha(T)$ curve such that northerly-dominant winds consistently magnify the attenuation coefficient at all wave periods.
Our analysis also suggests that attenuation rates are largest in magnitude in regime R2, which consists of mid-range values of the ice concentrations and mean wave periods, and relatively calm wave conditions.
This finding deviates from the results of \cite{rogers_etal21}, who found a positive (but weak) correlation between attenuation rates and ice concentration for the PIPERS data. 
These authors only used a subset of the data we analysed, however, as they only considered the east deployment (R1 and R3 for us). 
Comparing these two regimes is difficult for us as the $T$-range of validity of the power-law fits do not overlap. 
Extrapolating, however, does seem to suggest that a more concentrated ice cover attenuates waves more at a given period, which would be consistent with the conclusion of \cite{rogers_etal21}.

Although ice concentration is not sufficient to fully characterise the state of an ice field, our findings suggest a non-linear dependence of attenuation rates on ice concentration. 
Most large-scale modelling studies (typically done using WW3) parametrise ice-induced wave attenuation with a source term that is proportional to ice concentration \citep[see, e.g.,][]{liu_etal20}.
Our analysis suggests a less trivial picture, in which ice concentration, mean wave period and wind direction, at the very least, must be considered, with varying effects on different spectral bands.  
Therefore, we recommend that future empirical models of ice-induced wave attenuation should be piecewise and multi-parametric. 

\subsection{Buoys zonal spread and incident wave direction}\label{sec:dis-sprd}

In \textsection \ref{sec:an}.\ref{sec:an-attn}, we introduced an exclusion criterion for buoy pairs with relative angle $\theta_{rel}$ between the buoys and the incident wave direction exceeding $\theta_{max}=75^{\circ}$.
We chose this maximum angle to include a large number of attenuation coefficient measurements $\alpha$ in the dataset analysed in \textsection\ref{sec:res} while limiting the relative uncertainty on $\alpha$ caused by the uncertainty on the incident wave direction, which is assumed to be constant and directly south. 
We now evaluate the sensitivity of our findings on $\theta_{max}$. 

\begin{table}[h]
\label{tab4}
\begin{center}
\begin{tabular}{cc||c|c||c|c||c|c|}
   \multicolumn{2}{c}{}                                    & \multicolumn{2}{c}{R1}          & \multicolumn{2}{c}{R2}          & \multicolumn{2}{c}{R3} \\
                                               &           & $\tilde{v}\le0$ & $\tilde{v}>0$ & $\tilde{v}\le0$ & $\tilde{v}>0$ & $\tilde{v}\le0$ & $\tilde{v}>0$ \\     
    \hline
    \multirow{4}{*}{$\theta_{max}=60^{\circ}$} & $\beta$   & $3.2$           & $3.0$         & $1.9$           & $1.9$         & $1.5$           & $1.6$ \\
                                               & $\ln C$   & $-3.3$          & $-3.5$        & $-4.9$          & $-4.8$        & $-6.4$          & $-5.5$ \\
                                               & $T$ range & $[5,11]$        & $[5,10]$      & $[7,21]$        & $[10,21]$     & $[13,21]$       & $[12,21]$ \\
                                               & $N$       & $454$           & $1415$        & $9835$          & $6562$        & $29999$         & $71301$ \\
    \hline
    \multirow{4}{*}{$\theta_{max}=45^{\circ}$} & $\beta$   & XXX             & XXX           & $1.8$           & $1.7$         & $1.5$           & $1.6$ \\
                                               & $\ln C$   & XXX             & XXX           & $-5.4$          & $-5.4$        & $-6.4$          & $-5.7$ \\
                                               & $T$ range & XXX             & XXX           & $[10,21]$       & $[10,20]$     & $[12,21]$       & $[12,21]$ \\
                                               & $N$       & $0$             & $0$           & $6718$          & $3814$        & $24638$         & $44228$ \\
    \hline
    \multirow{4}{*}{$\theta_{max}=30^{\circ}$} & $\beta$   & XXX             & XXX           & $1.5$           & $1.6$         & $1.5$           & $1.5$ \\
                                               & $\ln C$   & XXX             & XXX           & $-6.3$          & $-6.0$        & $-6.7$          & $-6.2$ \\
                                               & $T$ range & XXX             & XXX           & $[9,21]$        & $[10,21]$     & $[12,21]$       & $[12,21]$ \\
                                               & $N$       & $0$             & $0$           & $3555$          & $1627$        & $19687$         & $26072$ 
\end{tabular}
\end{center}
\caption{Same as table~\ref{tab3}, but for three different values of the cutoff angle $\theta_{max}$.}
\end{table}

Table \ref{tab4} summarises the results of the power-law fits on $\alpha(T)$ (as was done in table \ref{tab3}) for the three cutoff angles $\theta_{max}=60^{\circ}, \,45^{\circ}$ and $30^{\circ}$.
The corresponding scatter plots in the  $(c,T_m)$-plane and $\alpha(T)$ curves are shown in the supplemental file (see figures~S6 and S7).
As expected the number of attenuation estimates decreases significantly with $\theta_{max}$. 
In fact for $\theta_{max}\le45^{\circ}$, no attenuation measurements in regime 1 were obtained.
Overall, decreasing $\theta_{max}$ has little effect on the slope $\beta$ of the power-law fit but it seems to cause a small downshift in the magnitude of the attenuation coefficient at all wave periods, similar to that caused by the wind direction. 

The downshift of $\alpha$ for decreasing $\theta_{max}$ can be explained by examining regime 2.
In that case, the parameter $\beta$ shifts from approximately 2 to 1.5 as $\theta_{max}$ decreases from $75^{\circ}$ to $30^{\circ}$. 
As this happens, attenuation coefficient measurements at the lower end of the $T_m$ and $c$ ranges in R2 are gradually filtered out, so that most attenuation estimates obtained for $\theta_{max}=30^{\circ}$ coincide in the $(c,T_m)$ plane with R3, for which ice concentration and mean wave period are higher.
The parameters of the power-law fit for R2 when $\theta_{max}=30^{\circ}$ are even very similar to those obtained for R3.
This suggests that decreasing $\theta_{max}$ tends to favour high-concentration ice conditions deeper into ice pack, i.e.\ where the buoys were originally deployed approximately along a north-south transect. 
There, short-period waves have already been filtered out (i.e.\ $T_m$ is relatively large) and wave attenuation rates are smaller.
This further reinforces our earlier finding that the magnitude of $\alpha(T)$ does not trivially depend on ice concentration, as $\alpha$ clearly shows a decreasing trend for high ice concentration.

The uncertainty on the relative angle between the incident wave direction and the buoys, denoted in \textsection\ref{sec:an}.\ref{sec:an-attn} by $\Delta\theta_{rel}$, is entirely dependent on that of the incident wave direction. 
As discussed earlier and in \cite{kohout_etal20}, wave direction was not recorded by the buoys during this deployment. 
We attempted to correct for this lack of information by analysing the hindcast WW3 wave data in the open ocean region just north of the ice edge. 
The incident directional spectrum was first approximated by using the average mean wave direction $\theta_{in}$ and average directional spread $\sigma_{in}$ in this region to construct the Gaussian distribution $\mathcal{N}(\theta_{in},\sigma_{in}^2)$.
Secondly, for each attenuation estimate, the incident wave direction was estimated by randomly sampling this Gaussian distribution and computing the effective separation between buoys along that direction $D_{i,j}$, accordingly.
Unfortunately, this procedure resulted in a majority of negative $\alpha$ values, i.e.\ wave growth events, which is not realistic.
Setting the incident wave direction to the mean wave direction extracted from WW3 data did not improve this issue and therefore motivated our choice to assume waves travelling on a north-to-south transect, for which the number of wave growth events is much more reasonable.

Our analysis suggests a severe limitation of conducting in-situ wave attenuation measurements in the MIZ with a large number of buoys and without recording wave direction.
Drifting buoys are highly likely to loose their post-deployment alignment along a north-south transect quickly, in which case the uncertainty on wave attenuation rate measurements increases quickly with the uncertainty on the incident wave direction.
This significantly limits the time window in which meaningful attenuation rates can be measured.
Future experimental programmes similar to PIPERS need to improve their capability in measuring wave direction.

\subsection{Other limitations}\label{sec:dis-lim}

In \textsection\ref{sec:res}, we conducted a correlation analysis of wave attenuation coefficients against wave, ice and wind drivers.
It is important to point out that the remote-sensing and hindcast data products used in this analysis have known biases and uncertainty, which could partly influence our conclusions. 
For instance, \cite{roach_etal18} showed that observational sea ice concentration products have significant variability depending on the algorithm used to analyze remote-sensing satellite data. 
This is especially true when concentration is high, i.e.\ $c>0.8$, which is the predominant ice regime observed during the PIPERS programme.
Clustering our wave attenuation estimates into the three regimes discussed in \textsection\ref{sec:dis}.\ref{sec:dis-proc}, where ice concentration ranges are relatively wide, partially limits the effect of this uncertainty.
WW3 wave data and ERA5 wind data are both based on model outputs, so that their reliability is limited by the fidelity of the physics used in the models. 
For instance, the version of WW3 that is used to produce open-water wave metrics near the ice edge does not model wave reflection by the ice edge.
Again, this uncertainty is unlikely to have a significant effect on our analysis, given that the bins used to group the independent variables, e.g.\ wind speed, wind direction, significant wave height, etc, are sufficiently wide to account for a large portion of the uncertainty.

Although our analysis has explored a large parametric space of drivers governing ocean wave attenuation in the MIZ, the effect of some quantities has not been considered. 
Most important to this study are ice thickness, floe size and ice type. 
It is well established that these properties of the ice cover are difficult to measure and there are no reliable remote-sensing data products easily accessible that would allow us to perform an analysis of the type done in \textsection\ref{sec:res}. 
There is evidence from both modelling \citep[e.g.,][]{montiel_etal16,boutin_etal18,meylan_etal21} and observational \citep[e.g.,][]{horvat_etal20,rogers_etal21} work that suggest sea ice properties play a key role in determining wave attenuation regimes.
Such properties were observed during the PIPERS programme at the time and place of each buoy deployment following the ASPeCt protocol. 
Although we have looked at this data, the sample size was too small and the ice conditions recorded were too homogeneous at deployment, so that no meaningful correlation analysis could be performed. 

\section{Conclusions}

We analysed ocean wave attenuation rates as a function of wave period collected in the autumn/winter Ross Sea marginal ice zone using in-situ wave buoys data during the 2017 PIPERS cruise.
Two deployments (four and nine buoys, respectively) at different times and locations were conducted.
This is the largest in-situ observational programme of this type to date in the Southern Ocean, during which a range of ice, wave and wind conditions were observed. 
This allows us to attempt formulating relationships between wave attenuation rates and its physical drivers.
Similarly to several recent observational studies, we found evidence of a consistent power-law relationship $\alpha = C T^{-\beta}$ between wave attenuation rates $\alpha$ and wave period $T$.
This relationship is limited in terms of spectral range, however, as a result of the high noise-to-signal ratio in the wave spectra at both short and long wave periods.
In these two extreme regimes, a spurious rollover effect in the $\alpha(T)$ relationship emerges, as demonstrated by \cite{thomson_etal21}.
After filtering out this effect, we attempted to quantify the extent to which ice, wave and wind physical drivers influence the $\alpha(T)$ power-law relationship.
Our main findings are summarised as follows.
\renewcommand{\labelenumi}{(\roman{enumi})}
\begin{enumerate}
    \item Three physical drivers emerged as key variables affecting the $\alpha(T)$ relationship, i.e.\ ice concentration $c$, mean wave period $T_m$ and wind direction $\tilde{v}$. 
    In particular, as either concentration or mean period increases, the power-law parameter $\beta$ decreases from approximately 3 to values less than 2.
    \item Beyond the expected positive correlation between ice concentration and mean wave period, three clusters of wave attenuation rates naturally appear in the $(c,T_m)$ plane, corresponding to three regimes of ice/wave conditions.
    \begin{enumerate}
    \item Regime 1 is characterised by low ice concentration, small mean wave periods and large waves (mean SWH $H_s>3$\,m), corresponding to a non-homogeneous ice cover in the MIZ within a few tens of kilometres of the ice edge, and mainly attenuates small-period waves (up to 10\,s) with power-law parameter $\beta\approx3$. 
    This is the low-pass filtering effect of the MIZ often discussed in the literature and is likely to be mostly due to wave scattering ice floes.
    \item Regime 2 is distinguished by mid-range ice concentration and mean wave period values, and calm wave conditions with average SWH $H_s\approx0.5$\,m.
    In this regime, waves attenuate over a broad spectral range with attenuation rates governed by a power-law with power $\beta\approx2$.
    It is likely that both dissipative and scattering processes cause the observed attenuation in this regime.
    \item Regime 3 corresponds to ice conditions with very high concentration, deeper into the MIZ ($d_{\mathrm{edge}}>100$\,km), likely to be nearly continuous ice.
    Wave conditions are very calm (with SWH $H_s<0.5$), with spectra mainly composed of long-period waves ($T>10$\,s), so that all shorter-period have been filtered out closer to the ice edge.
    In this regime the power-law describing attenuation rates has parameter $\beta<2$, and dissipative processes are likely to be the dominant effects causing the attenuation.
    \end{enumerate}
    \item Wind direction affects the magnitude of the attenuation coefficient, as southerly-dominant winds consistently shift upward $\alpha$ values at all wave periods (i.e.\ $C$ increases), while northerly-dominant winds consistently shifts those values downward (i.e.\ $C$ decreases). 
\end{enumerate}

Our findings suggest that simple, catch-all parametrisations of the attenuation coefficient $\alpha$ in spectral wave models, e.g.\ WW3, are unlikely to perform well given the wide range of wave, ice and wind conditions encountered in the MIZ, especially in the Southern Ocean.
Remote sensing estimates of attenuation rates captured over two years in the Southern Ocean reveal $\alpha$ can span 3--4 orders of magnitude \citep{stopa_etal18}, suggesting a strong influence of multiple oceanic, ice and atmospheric physical drivers. 
Future model developments should take into account these dependencies and acknowledge the large uncertainty in model predictions given the large uncertainty in the physical drivers causing the observed ice-induced wave attenuation.

Our analysis has also demonstrated important limitations in extracting attenuation data from pairs of wave buoy spectra.
Most important is the lack of knowledge of wave directionality, which generates significant uncertainty in the estimated attenuation coefficient.
Although the model inversion approach used by \cite{rogers_etal16,rogers_etal21} partly circumvents this issue as the forcing directional wave field is simulated directly from wind inputs, it brings up different and not less important issues.
For instance, a constant attenuation rate is assumed between the ice edge and the wave buoy analysed, therefore not accounting for the potential variability in ice, wave and wind conditions in the system between these points, apart from a linear scaling by ice concentration, which we have shown here does not seem to be valid.
More importantly, the model itself makes a number of assumptions about how source terms are parametrised in the MIZ, which has not been validated.
In conclusion, it is clear that the exercise of parametrising ice-induced attenuation is far from being settled. 
Much more work is needed to understand the different processes involved and the conditions in which they dominate.

\clearpage
\textit{Acknowledgements.}\quad
We thank Steve Ackley and the captain and crew of {\it RV Nathaniel B. Palmer} for their assistance in deploying the waves-in-ice instruments.
The authors acknowledge the preliminary analysis of the PIPERS conducted by Timo Milne (summer student working with FM).
The work was funded through New Zealand’s Deep South National Science Challenge \emph{Targeted Observation and Process-Informed Modelling of Antarctic Sea Ice} (contract C01X1445). 
FM also acknowledges financial support from Royal Society Te Ap\={a}rangi (Marsden Fund projects 18-UOO-216 and 20-UOO-173) and the New Zealand's Antarctic Science Platform (Project 4). 

%
%
\textit{Data availability statement.}\quad
The PIPERS wave-ice data is available upon request from  Alison Kohout (alison.kohout@niwa.co.nz).
WAVEWATCH III\textsuperscript{\textregistered} ocean wave hindcast data are freely available for download on the NOAA/NCEP website \url{https://polar.ncep.noaa.gov/waves/download.shtml?}. 
Passive microwave sea ice concentration data are freely available for download on the NSIDC website \url{https://nsidc.org/data/G02202/versions/3}. 
ERA5 reanalysis wind data are freely available for download on the Copernicus Climate Change Service (C3S) Climate Data Store (CDS) website \url{https://cds.climate.copernicus.eu/cdsapp#!/dataset/reanalysis-era5-pressure-levels?tab=overview}.


\bibliographystyle{apalike}

\bibliography{references}

\end{document}